# Anomalous electric-field effect and glassy behaviour in granular aluminium thin films: electron glass?

T. Grenet[a], J. Delahaye, M. Sabra, F. Gay

*LEPES-CNRS, BP 166, 38042 Grenoble cedex 9, France*
*Fax: 19 (0)4 76 88 79 88*
*Tel: 19 (0)4 76 88 74 61*

**Short title**: Glassy behaviour in granular aluminium thin films

**Abstract.** We present a study of non-equilibrium phenomena observed in the electrical conductance of insulating granular aluminium thin films. An anomalous field effect and its slow relaxation are studied in some detail. The phenomenology is very similar to the one already observed in indium oxide. The origin of the phenomena is discussed. In granular systems, the present experiments can naturally be interpreted along two different lines. One relies on a slow polarisation in the dielectric surrounding the metallic islands. The other one relies on a purely electronic mechanism: the formation of an electron Coulomb glass in the granular metal. More selective experiments and/or quantitative predictions about the Coulomb glass properties are still needed to definitely distinguish between the two scenarii.

**1. Introduction**

We present here an experimental study of out of equilibrium phenomena observable in the electrical conductivity of insulating granular aluminium thin films. We focus on an anomalous electric field effect and on its glassy relaxation subsequent to external disturbances, brief accounts of which were previously given in [1, 2]. In this section we first give a short overview of the experimental context of our studies.

---
[a] Corresponding author

As far as we know the phenomena of concern here were first observed and reported in thin granular metal films (granular gold) [3]. The authors performed field effect measurements at low temperature, using MOS like structures, which consist in the superposition of a metallic gate, a dielectric spacer and a granular metal film (the channel) which conductance is investigated. For insulating granular channels, they found a symmetrical minimum in the conductance versus gate voltage curves ($G(V_g)$), whose position was shown to be determined by the applied $V_g$ while cooling the sample. The finding was interpreted in terms of Coulomb blockade of small grains and "potential disorder" due to charges trapped in the substrate (offset charges in today's language). In brief, the absence of a true macroscopic Coulomb blockade was ascribed to the existence of the "potential disorder", whose role had been overlooked in previous discussions of the electrical transport in granular metals. The very existence of the observed $G(V_g)$ minimum, and of its memory effect, was ascribed to a slow dynamics of these trapped charges.

Later a similar field effect was found in indium oxide ($InO_x$) insulating thin films [4], and has been extensively studied since then [5]. In particular the slow relaxation of the field effect minimum, induced by disturbances like gate field or temperature abrupt changes, was studied in detail. The phenomena were interpreted from a different point of view than in [3]. The authors argued they are the signature of a correlated electron glass, resulting from the strong Coulomb interactions between localised carriers (Coulomb glass), which had been theoretically proposed earlier [6]. The effect of the electron doping level on the $G(V_g)$ minimum and its slow dynamics was shown to be consistent with the Coulomb glass interpretation [7].

Qualitatively similar effects were also reported in a system consisting of an ultra-thin lead film deposited on an amorphous Ge wetting layer [8, 9]. Very recently slow relaxation phenomena were reported in silicon MOSFETs [10].

As far as we know these experiments are the only ones to demonstrate macroscopically long times (more than hours or days) which may indicate the existence of a Coulomb glass. While experiments on $InO_x$ have been extensive, much less is known about the other case reported (granular/ultra-thin metal films), although its investigation may help to progress in the understanding of the mechanisms at work. Our current study of granular Al thin films is aimed at filling this gap.

The next section is devoted to the choice of the system and the samples preparation and characterization. We then describe the out of equilibrium phenomena observed in the electrical conductance (section 3). Finally, the results are compared with those of $InO_x$, and are discussed in the light of two different possible mechanisms and of recent theoretical works.

## 2. Granular aluminium films, samples preparation and characterisation

In order to test the universality of the out of equilibrium phenomena studied in $InO_x$, we searched for them in systems a priori quite different from the structural point of view, but still having similar electronic properties. We chose to study granular metals, which display a metal to insulator transition and transport properties analogous to other disordered systems. Generally (but not exclusively), their conductance follows a $\exp(-(T_0/T)^{1/2})$ law up to room temperature (or even above). While the reason for this has been the subject of a long debate [11], there seems now to be a consensus that it corresponds to a variable range hopping in the presence of strong Coulomb interactions [12].

### 2.1 Sample preparation:

We focused on granular aluminium, which consists of small metallic Al islands separated by insulating $Al_2O_3$ barriers. In principle one may study how the physical properties depend on the micro-structural characteristics (e.g. $Al_2O_3$ barrier thickness and Al islands sizes) which can be useful to elucidate the microscopic mechanisms at work. Moreover granular aluminium samples are quite stable, as compared for instance to UHV stored ultra-thin metals or granular gold films, whose connectivity generally evolves after their preparation, due to the free surface atomic diffusion. In our case, granular Al films can be repeatedly cycled between room temperature and 4 K for weeks, without any significant change of the low temperature conductance values. The samples were fabricated by e-beam evaporation of Al in controlled oxygen pressures ($P_{O2}$). Obviously the metallic or insulating character of the films is determined by the ratio $P_{O2}$ / Al evaporation rate. Typically, the films we focus on here were obtained with $P_{O2} \sim 10^{-5}$ mbar and evaporation rates around 2 Å/sec. The films were deposited either on sapphire substrates or on oxidized doped silicon wafers, at room temperature. In the first case, for the field effect measurements, an Al gate and an alumina gate insulator (1000 Å thick) were evaporated prior to the granular Al channel, and in the last step source and drain Al contacts were evaporated on the granular Al channel. All these steps were performed without opening the chamber. On wafers the thermally grown $SiO_2$ and the conducting doped Si were used for the gating, gold contacts were deposited on the oxide prior to the granular Al film. In order to reduce the measured resistances, wide and short channels were mostly used (typically 1.8 mm x 20 µm).

### 2.2 Microstructure:

Granular Al is known to consist of Al islands separated by $Al_2O_3$ barriers [13]. While Al evaporation in a good vacuum produces large Al grains (even larger than the films thickness), evaporation in a given $P_{O2}$ results in smaller Al grains, whose growth is stopped by the formation of an oxide layer on their surface, on which secondary growth of Al islands then occurs [14]. The higher the $P_{O2}$, the smaller the mean grain size and the thicker the average $Al_2O_3$ barriers. Most of our films have a thickness of 200 Å. We noted in [1, 2] that AFM as well as scanning electron microscopy pictures of our samples show "islands" of size around 100-200 Å. However, X-ray diffraction (XDR) measurements, performed on thicker films (5000 Å) in order to increase the signal, indicate smaller grain sizes. In Figure 1 we show the Al (111) and (200) diffraction lines for a series of samples evaporated under different $P_{O2}$. One observes that while the lines remain centred on the same angles, they broaden considerably for samples close to or in the insulating state, indicating a dramatic reduction of the grain size. We may estimate the grain sizes using the Scherrer formula, which reads in our case: Size (Å) ≈ 1.5 / $\Delta\theta$, where $\Delta\theta$ is the FWHM of the peaks measured in radians. For samples (a) to (c) we thus estimate respectively 1200 Å, 120 and 90 Å, and guess a smaller value for the insulating sample (d). Note that these are volume weighted average sizes, so that number weighted average sizes should be smaller (by a factor determined by the grain size distribution, which we unfortunately cannot estimate from these data). We can anyway conclude that the grain sizes of the insulating samples we are interested in here, are of a few nanometers, say typically 5 nm. This estimate is in accordance with the results of other studies [13, 15]. The Al island size is primarily determined by the oxygen pressure (and evaporation rate), and was reported to be independent of the substrate nature and film thickness [15]. Strictly speaking our 200 Å thick samples are thus not two dimensional arrays of islands, as the films thickness equals several times the island size. These sizes we are estimating are the "crystallographic" sizes of the Al grains, however one would like to know what the "electrical" sizes are, which are determined by the grains or clusters of grains which are separated from the rest of the sample by a barrier resistance higher than $R_Q$ ~ 26kΩ. This is of importance when discussing the electrostatic charging energies involved in electron transfers. Unfortunately we cannot determine these sizes experimentally in any simple way. Since the grain growth during Al deposition is limited by the encapsulation with an oxide layer, the "crystallographic" sizes may be a reasonable estimate. Assuming "electrical" islands of a few nanometers (and $Al_2O_3$ barriers ~ 10 Å thick) the electrostatic charging energies are larger than most of our working temperatures.

**2.3 Electrical properties:**

The $R(T)$ curves of a series of 200 Å samples is shown in Figure 2. Most of the experiments to be discussed in section 3 were performed on insulating samples with $R(4K)/R(300K)$ ~ $10^3$–$10^5$, which offer a compromise for not too high resistances at 4 K and still easily measurable anomalous field effects. As already noted in [1] the divergence of resistance as $T \rightarrow 0$ is faster than the $\exp((T_0/T)^{1/2})$ law generally reported for granular metals and cermets (this finding will be discussed elsewhere). We note that the out of equilibrium effects of interest here do not seem to depend on the precise $R(T)$, they are equally observed in $InO_x$ films obeying either simple activation or variable range hopping dependences [16].

As illustrated in the inset of Figure 2, the samples are easily driven in a highly non ohmic regime, but when the bias electrical field is below ~ 5 $10^3$ V/m (at 4 K), the samples are ohmic. The conduction mechanism in granular metals is thought to result from an interplay between Coulomb blockade, disorder and thermal activation. In our samples, no sign of a macroscopic Coulomb blockade was observed (which we expect as a non ohmicity at low $E_{bias}$), showing that disorder plays an important role. Unless otherwise specified, all our measurements were performed in the ohmic regime.

## 3. Anomalous field effect and non equilibrium phenomena

In Figure 3 we show a typical $G(V_g)$ curve obtained at 4 K with a MOS structure whose granular Al channel has $R(4K)/R(300K) = 2.10^3$ and $R_\square = 32$ MΩ. The sample was cooled with $V_g = 0$ and kept for one day at 4 K before the measurement was performed. One clearly sees the "cusp like" symmetrical conductance minimum, whose existence and properties are the subject of this paper (in the following this anomaly will be referred to as the "cusp" or the "dip"). The cusp, as measured at 4 K, typically has an amplitude of a few percent, and a full width at half maximum (FWHM) $\Delta Vg_{1/2}$ of 1 V. We may express the cusp FWHM as the surface charge density $Q$ induced by a gate voltage equal to $\Delta Vg_{1/2}$ (using $Q=C\Delta Vg_{1/2}$ with in our case $C=\varepsilon_0\varepsilon_r(Al_2O_3)S/t$, $\varepsilon_r(Al_2O_3)=9$, $S=1$cm$^2$, $t=10^3$ Å). This parameter does not depend on the gate insulator nature and thickness and allows to compare different systems. We find $Q = 5.10^{11}$ e$^-$/cm$^2$ (1 e$^-$ / (140x140) Å$^2$). The cusp was reproducibly observed in all insulating granular Al films studied (tens of samples), using various measurement configurations. In particular, we measured samples deposited on sapphire or oxidized silicon, with DC or low frequency AC biases (below 100 Hz), involving two contacts configurations (using a $V_{bias}$ source and a current to voltage converter to measure the current) or four contacts configurations (with several contact geometries, using an $I_{bias}$ source and a nanovoltmeter to measure the voltage). The cusp cannot be caused by current leaks through the gate insulator, since these were generally directly checked to be too small to cause the observed effect. Moreover a current leak is easily detected as it gives rise to a DC antisymetric component to the $G(V_g)$ curve (as was observed in a few leaky structures). In AC measurements any current leak is efficiently eliminated by the lock-in amplifier, so that identical DC and AC $G(V_g)$ curves show that current leaks are of no importance and that the cusp reflects a property of the channel's conductance.

The amplitude of the cusp depends on the measurement duration [2]: it increases logarithmically with the $V_g$ scan rate used to record it. This shows that its existence is related to non equilibrium phenomena. However, as shown in [2], the shape and width of the cusp do not depend on the measurement duration.

### 3.1 "Steady" properties of the cusp:

Strictly speaking the samples are never in their equilibrium state. However as the relaxations to the equilibrium state are logarithmic in time, once a sample has been allowed to relax for a sufficient time (say one or a few days), the field effect curve evolves very slowly and does not change significantly during a typical measurement time. We can thus define and study "steady" properties of the cusp, and observe the following features:

- the cusp depends on the equilibration temperature: it is prominent at low $T$ but is washed out at high $T$. More precisely, as shown in Figure 4, its width increases linearly with the equilibration $T$ while its *relative* amplitude sharply decreases (like $1/T^2$ in the example shown). Note that owing to the conductance temperature dependence, the cusp *absolute* amplitude increases with $T$ in the temperature range investigated, above which it saturates and then presumably diminishes. We note that the numerous samples equilibrated at 4 K all displayed the same cusp width, independently of their $R_\square$, dimensions or substrate. The cusp width is thus determined by the temperature [17] (of course when comparing samples with different gate insulators, the cusp width has to be expressed in induced charge density).

- the cusp relative amplitude depends on the sample's insulating strength: the higher the $R_\square$ (say at 4 K), the more prominent the cusp, as illustrated in Figure 5.

- the cusp is also altered by the bias voltage when the latter is high enough to drive the sample into the non ohmic regime. Qualitatively, we observed that an increase of the (non ohmic) bias is equivalent to an increase of the sample's temperature, although the latter is maintained at 4 K. When $V_{bias}$ is increased, the conductance increases (non ohmicity), the cusp broadens and its relative amplitude diminishes like if the temperature had risen [18].

**3.2 Slow relaxation of the cusp**

3.2.1 Temperature induced relaxation

As mentioned above, strictly speaking, the samples are never in equilibrium. Indeed after cooling from room temperature to say 4 K, keeping $V_g$ constant, one always observes a logarithmic decrease of the sample's conductance, without any saturation even after several days. It is instructive to monitor the $G(V_g)$ cusp during that relaxation, as shown in Figure 6-a. One observes a slow evolution of the cusp, from a smaller and rounder dip, towards the "steady" 4 K shape and amplitude. During that relaxation, the cusp amplitude increases logarithmically with time, as shown in Figure 6-b, which explains the relaxation of the conductance measured at constant $V_g$. The sample thus keeps some memory of its thermal history: looking at the series of curves in Figure 6-a, one can deduce that the sample, now at 4 K, was previously maintained at a higher temperature. But this memory is not permanently frozen and is progressively lost. We insist on a striking feature of the upper curve: the sample has its 4 K conductance which means that its electrons essentially do conduct like in a sample equilibrated at 4 K, but the cusp still has a rounded shape, reminiscent of a temperature when the sample's conductance was significantly larger.

Note that the slow logarithmic decrease of the conductance after cooling is also measured on samples without a gate. In that case the granular Al film is deposited directly on its single crystalline sapphire substrate. This shows that the slow relaxation is a property of the granular Al film, and not of the disordered alumina or silicon dioxide insulators of the gated samples.

### 3.2.2 Gate voltage induced relaxation

Another way to display a memory effect and induce a slow relaxation of the cusp is to change the gate voltage and maintain it at the new value, like in the "two dip" experiment. What happens is shown in Figure 7. Initially equilibrated at 4 K with $V_g = 0$ (step 1), the sample has a well developed cusp as shown in the upper curve, corresponding to a "steady state". The gate voltage is then set at $V_g = 2.5$ V for $t_w = 3$ hours, which provokes a progressive erasure of the initial cusp and creates a new one centred on $V_g = 2.5$ V (step 2). In step 3, the gate voltage is set back to zero, the second dip disappears while the first one recovers. Thus when one imposes a given gate voltage, a cusp centred on that voltage value is progressively written, while all other anomalies are progressively erased.

This observation explains the above mentioned dependence of the cusp amplitude on the $V_g$ scan rate, and in the limit of an infinitely slow scan, the cusp has "enough time" to "follow the gate voltage", so that the measured $G(V_g)$ curve is flat. Note that the relaxations we observe cannot be related to a slow "macroscopic" charging of the granular channel once $V_g$ was changed. First the observed time scales are much longer than the $RC$s of the measuring circuits, and than the estimated Maxwell times of the granular Al films (which give the time it takes a macroscopic charge to spread on the film). Second, a slow evolution of the dielectric constant of the gate insulator, due to a slow evolution of its polarization after a $V_g$ change, would presumably cause a continuous slow drift of the cusp position in the $G(V_g)$ curve, but not the "two dip" pattern we observe.

### 3.2.3 Aging

Aging arises from the fact that the response of a system to a given stimulus depends on what external conditions were imposed to it previously, and for what duration. In Figure 8, we illustrate the kind of aging our system displays in step 3 of the "two dip experiment" (TDE). In Figure 8-a we show the time dependent amplitudes of the second cusp while it is erased (decreasing curve) and of the first one while it is restored (increasing curve) [19]. It is seen that the two cusps evolution during step 3 are symmetrical, the sum of their amplitudes being constant (middle curve). In Figure 8-b, the TDE was conducted with different durations $t_w$ of step 2 (writing of the second cusp). We show the decreasing amplitude of the second cusp during step 3, for different writing durations ($t_w$), and note that the different vanishing amplitude curves are simply shifted on a log scale, their logarithmic part being parallel, and that they perfectly collapse on a single master curve when plotted as a function of $t/t_w$, corresponding to what is called "simple aging".

One can also study aging effects in the step 2 of the TDE (instead of step 3), which is a more usual procedure: $t_w$ is now the time elapsed between the cooling of the sample from room temperature and the application of the perturbation (sudden change of the gate voltage). As can be seen in Figure 9, we observe that the two cusps amplitudes first evolve like Ln($t$) with opposite slopes and that when the first cusp is essentially erased the second one grows without saturation. If the experiment is repeated with a larger $t_w$, the first cusp naturally has a higher amplitude and its decreasing curve is simply shifted along the Ln($t$) axis (corresponding again to simple aging). But the second cusp amplitude grows independently of $t_w$: no aging is observed for it.

In summary, we observe that the erasure of a cusp written for a given duration $t_w$ obeys simple aging and lasts typically during $t_w$. The writing of a new cusp proceeds logarithmically with time without any sign of saturation, and does not depend on the previous history. When logarithmic the amplitude evolutions all proceed at the same rate (for a given sample).

### 3.2.4 Relaxation at different temperatures

One may wonder what the effect of the temperature on the $V_g$ induced slow dynamics can be. One simple way to tackle this question may be to repeat the same TDE experiment at different temperatures. The result of such an experiment is displayed in Figure 10. There we show the decreasing amplitude of the second cusp once $V_g$ was set back to its initial value (step 3 of the TDE). One sees that apart from the temperature dependent amplitude of the cusp, the relaxations all follow the same "master curve", at least up to 20 K. The vanishing amplitude makes it difficult to perform accurate measurements at higher temperatures. This experimental result shows that the system still has macroscopically long times (at least $10^5$ sec) at 20 K. But it does not mean that the relaxation times involved are temperature independent. To illustrate this let us imagine that during the TDE, the two cusps evolution is due to the slow dynamics of some independent "degrees of freedom" of the system evolving "forth" (step 2) and then "back" (step 3), each with their own characteristic relaxation time $\tau_i$. If these "individual" relaxation times depend on temperature ($\tau_i(T)$), then at each temperature change the $\tau_i$ distribution is shifted and the experimental time window (from a few to $10^5$ secs in our example) corresponds to a different section of it. If the Ln($\tau_i$) distribution is wide enough and flat then the experimental master curve is not expected to depend on temperature. The $T$ dependence of the $\tau_i$ is not seen as long as one of the tails of the $\tau_i$ distribution hasn't entered the experimental time window.

A more stringent search for a possible $T$ dependence of the relaxation times is then the following: perform a series of TDE experiment with step 2 and step 3 performed at different temperatures; for example performing step 2 with the same $T$ (and of course the same $t_w$), and performing step 3 at different temperatures. In that case if the *same* relaxation processes are involved "back" and "forth" in steps 2 and 3, one should observe different relaxation curves for step 3. We show the result of such a comparison between two temperatures in Figure 11. The second cusp was written at 5 K in both cases (step 2), while it was erased (step 3) at 5 K in one run, and at 8 K in another run. Clearly, apart from the amplitude normalisation, both curves are identical, which shows that the erasure of a cusp is not accelerated even if performed at a higher temperature than its writing. We conclude that, at least in this limited temperature range, if the $V_g$ induced cusp dynamics is due to a collection of degrees of freedom which reversibly relax, then these are not thermally activated. These results should of course be pursued at higher temperatures.

## 4. Discussion

We first summarize the experimental facts described above. We have shown that the conductance of insulating granular aluminium thin films relaxes logarithmically to a lower value once the samples have been cooled from room temperature. The relaxation corresponds to the slow formation of a sharp minimum in the $G(V_g)$ curve centred on the cooling $V_g$, which is most prominent at low temperature and in highly insulating samples. The application of a non ohmic bias voltage alters that "cusp" similarly to an increase of temperature. Changing the gate voltage erases the existing minimum and writes a new one centred on the new $V_g$. The relaxations obey simple aging and are logarithmic with time as long as a finite time history does not interfere. The gate voltage induced relaxation rates are not thermally activated (at least at low temperature).

### 4.1 Comparison with the indium oxide phenomenology

As we noted in the introduction, qualitatively similar phenomena have been reported in a few other systems. We will restrict the comparison of our results to those obtained with indium oxide films, which have been studied extensively, reported in great detail, and have inspired our studies. The main point is the striking qualitative similarity between the behaviours of granular aluminium and indium oxide. All the qualitative features which we have described here have also been observed in the latter [5].

The similarity may also be quantitative. The cusps amplitudes are similar, and the widths are also comparable (at a given $T$). More precisely, in $InO_x$ the cusp widths depend on the oxygen concentration (electron doping level). When expressed in gate induced surface charge density, they are the same in strongly doped $InO_x$ films as in granular aluminium. The TDE master curves (Figure. 8-b) are also very close.

Recently aging experiments equivalent to the one we show in Figure 9 have been reported for indium oxide [20], and a simple aging has been discussed, associated with the growth of the second cusp. This seems to contradict our observation that the second cusp amplitude grows independently of $t_w$. Actually the discrepancy is only apparent. In [20] the aging appears because the quantity discussed is not the second cusp amplitude (which we evaluate as $G_{base\ line} - G_{minimum\ of\ second\ cusp}$) but a quantity which combines both the first and second cusp amplitudes (the difference between the actually measured conductance and the one that would be measured had $V_g$ not been changed).

Some real differences may however be pointed out, but we feel these still need more experimental inspection. In InO$_x$, the magnetic field was reported to slow the $V_g$ induced relaxation [21], although it has no effect on the "steady" cusp itself. In our case, no alteration of the steady cusp or of the dynamics of the TDE was observed up to 7 Tesla at 4 K. For example, in the course a TDE, switching on or off the magnetic field has no effect on the dynamics observed. As for the effect of the temperature, recent studies in InO$_x$ suggested a deviation of the $T$ dependence of the cusp width from the linear behaviour at low $T$ [16], and an increase of the cusp dynamics when lowering the temperature [22]. Future experimental efforts will be devoted to the investigation of these differences.

### 4.2 Relaxation functions

The Ln($t$) dependences we observe for the conductance or cusp amplitude time evolutions, suggest to analyse the relaxations as resulting from the sum of individual contributions. Let us suppose that we have a collection of "degrees of freedom" in the system, which respond slowly to external disturbances. For instance in the TDE, they may relax "forth" when the gate voltage is changed to a new value, and then "back" to their initial state once $V_g$ is set back to its initial value. Let $\tau_i$ be the relaxation time of a degree of freedom "$i$", meaning that if a disturbance is applied at time $t = 0$ the probability it has relaxed in response to it at time "$t$" is $(1-\exp(-\frac{t}{\tau_i}))$. Then during step 2 of the TDE the conductance has a variation (second "dip" amplitude) given by:

$$\Delta G(t < t_w) = \Delta G_0 \sum_i (1-\exp(-\frac{t}{\tau_i})) \qquad (1)$$

At $t = t_w$ the process is reversed (step 3, $V_g$ set back to its initial value). If we further assume that the relaxations "back" and "forth" have the same relaxation times, the second "dip" amplitude is erased like:

$$\Delta G(t > t_w) = \Delta G_0 \sum_i (1-\exp(-\frac{t_w}{\tau_i}))\exp(-\frac{t-t_w}{t_i}) \qquad (2)$$

With relaxation times such that Ln($\tau_i$) has a flat distribution, these formula reproduce naturally the experiments. They indeed give: i) simple aging, ii) logarithmic relaxations in step 2 or in step 3 for $t<t_w$, iii) a saturation in step 3 ($t \geq t_w$), with a calculated relaxation function very close to the master curve experimentally determined as shown in Figure 12. We stress that, unlike other approaches involving stretched exponentials or trap models [20], the remarkable agreement between the calculated and measured master curves is obtained here without any fitting parameter. Thus the macroscopic conductance relaxations we observe can be well described as the sums of distributed independent relaxations (the same is true for the resistance, owing to the small relative amplitude of the changes).

Different master functions or aging laws can of course be obtained with such a model, depending on the relation between the forth and back relaxation times ($\tau_{i\rightarrow}$ and $\tau_{i\leftarrow}$). For instance with $\tau_{i\leftarrow} = k\tau_{i\rightarrow}$ one still gets simple aging but with a master curve saturating for $\frac{t}{t_w} \approx k$ (instead of $\frac{t}{t_w} \approx 1$), and for $\tau_{i \leftarrow} = (\tau_{i \rightarrow})^\mu$ one gets a master curve scaling with $\frac{t}{(t_w)^\mu}$. Although the precise mechanism at work is not known, one can imagine that upon increasing the amplitude of the perturbation applied to the system to induce a relaxation, the relation between $\tau_{i \rightarrow}$ and $\tau_{i \leftarrow}$ can be modified and the simple aging can be lost, as was observed in TDE experiments with InO$_x$ for large enough $V_g$ changes [23]. One such mechanism could involve the modification of the potential profile of some tunnelling entity by the applied electric field (see below).

### 4.3 Do we have an electron Coulomb glass?

Our description of the relaxation curves is purely phenomenological and tells nothing about the physical origin of the system's slowness. In the remaining of this paper we want to discuss different scenarii one can envisage.

Since we could not find any signature of a thermal activation of the dynamics, we suppose it proceeds by tunnelling events. The tunnelling entities may be ions or electrons. If the tunnelling events are individual, a quick estimate shows that for ions the tunnel barriers must be rather low (in the meV range), otherwise the relaxation rates would be un-measurably long. For electrons on the contrary the relaxation times would be un-measurably fast unless the tunnel barriers are rather thick (several tens of angstroms). Another possibility to have long enough electronic relaxation times is that the electrons are highly correlated and each tunnelling event involves many electrons.

In the case of ions, one can imagine that the conductance relaxations we observe are due to the influence of slowly relaxing ionic configurations on the electron conducting channels. We call this the *extrinsic scenario*. In the case of individual electron tunnelling, it is hard to believe that the conduction proceeds only by such low frequency hops, because then the samples resistance would be un-measurably high. It is more reasonable to assume that conduction proceeds by easier electron tunnelling events, and that the slow localised electron events influence the conduction channels, just like the ions do in the extrinsic scenario. Finally the highly correlated electron scenario corresponds to the *Coulomb glass scenario*.

We now discuss the scenarii in more detail, and try to see to what extent they can explain the field effect measurements.

### 4.3.1 The model of Cavicchi and Silsbee

A kind of extrinsic scenario was first advocated by Adkins et al. [3] to explain the occurrence of a $G(V_g)$ cusp in granular gold films. Actually this explanation, like the experimental results, is very reminiscent of findings first reported forty years ago [24], which were reproduced and discussed in more details later by Cavicchi et al. [25]. We briefly recall these early works and the essence of the model proposed, more details are given in the appendix. For simplicity we use the notations of Cavicchi et al.

The experiments described in [24] and [25] consist in measuring the differential capacitance $C$ of a plane capacitor which contains small isolated metallic islands inside its insulator, close enough to one of its two plates so that electrons can tunnel back and forth between that plate and the islands. In today's terminology the islands constitute a parallel array of "electron boxes". Measurements of the differential capacitance of that system as a function of an applied DC voltage $V_S$ revealed a cusp, with a phenomenology very similar to that of our $G(V_g)$ cusp (bias induced slow relaxation and memory effects (TDE), thermal memory). The analysis of the phenomena was as follows (see the appendix for more details).

The tunnelling events between the islands and the nearby plate contribute to the global capacitance measured. At low temperature their occurrence is governed by Coulomb blockade. More precisely one would expect each island to provide a periodic contribution to $C$ measured as a function of $V_s$. It is tempting to seek for the origin of the $C(V_s)$ anomaly in these Coulomb blockade oscillations, however the authors showed that due to the ever present island potential disorder, these should be dephased and no signature should remain in $C(V_s)$. A mechanism for the origin of the cusp and its slow dynamics was proposed: a slow relaxation of the polarisation of the dielectric surrounding the metallic islands. In brief, the interplay of that polarisation relaxation and of the discrete changes of island charge states it induces, partially rephases the individual periodic contributions, resulting in the formation of a cusp centred on the waiting $V_s$. This mechanism has some similarity with the one encountered in e.g. spin glasses. For a given $V_s$, the islands can be divided into "soft" (their charge state easily fluctuates) and "hard" ones. The slow formation of the $C(V_s)$ cusp is due to a slow evolution of the local electric potentials (or fields) which reduces the number of "soft" islands. However that mechanism is "extrinsic" because the slow cusp evolution reflects the dynamics of the dielectric polarisation surrounding the islands, and is not intrinsic to the charges on the islands.

The origin of the slow polarisation is not precisely known. But since several decades, a wide variety of experiments have demonstrated the ubiquity of slow dynamics in a wide range of materials at low temperature. The anomalous thermal and acoustic properties of amorphous solids at low $T$ (below ~1 K) have been attributed to atomic or ionic two level systems (TLS), which relax with macroscopic characteristic times [26]. These have been observed in metallic or insulating amorphous materials, as well as in polycrystalline materials. The low frequency noise observed in small semiconductor devices [27] or in planar tunnel junctions [28] at low temperature is believed to be caused by charge carrier trapping/detrapping in the dielectric. The slow dynamics of this process was attributed to slow atomic/ionic TLS which jumps alter the traps configurations. Signatures of atomic jumps at low temperature have also been observed in mesoscopic quantum interference effects [29] and in point contact spectroscopy [30]. A slow relaxation of the polarisation of amorphous dielectrics at low temperatures has been studied in detail [31], also attributed to the dynamics of ionic TLS. Interestingly, a cusp was observed in the $C(V)$ curves, showing a dynamic phenomenology quite similar to the one we presented above. In that case, the slowness of the relaxation is believed to be intrinsic to the isolated TLS, but the existence of the cusp is ascribed to interactions between them. This case is thus intermediate between the "extrinsic" interpretation of the cusp given by Cavicchi et al. (involving a collection of independent entities) and the Coulomb glass interpretation discussed below (for which the slow dynamics *and* the cusp are attributable to the long range interaction between the carriers localized on the islands). In closer connection to our samples, the effect of "offset charges" on the properties of Coulomb blockade devices was outlined since their beginning [32]. It has been studied in some detail more recently [33] and discussed in terms of TLS present in the oxide surrounding the metallic islands.

To summarize, there is ample evidence that at low temperature, even though long range atomic or ionic diffusion may be absent, local jumps are always present and influence the electronic properties in various ways. Actually the picture of a two level system is oversimplified and there is experimental evidence that clusters of atoms/ions (maybe composed of tens of them), possessing several intermediate metastable configurations, are present [34]. The dynamics of such entities may be complex. Their relaxation can be induced by temperature, voltage or light illumination. Although at high temperature the relaxation rates are generally thermally activated, at low temperature they may be temperature independent or, in some cases, increase when *T* is decreased [35]. It was even shown that the current flowing through the devices can influence the dynamics of the TLS [33].

We believe that these facts cannot be ignored when considering our results, and that the purely electronic nature of the glassy behaviour we observe is not self evident. In the next section we try to see to what extent the "extrinsic" interpretation may or may not explain our findings.

**4.3.2 Application of the extrinsic scenario to our experiments**

The experiments discussed in [24] and [25] are simpler than ours because the metallic islands are measured in parallel so their contributions to the capacitance are additive, and the islands are assumed to be independent from each other (no electrostatic coupling nor charge transfer from island to island). On the contrary in our case the measurement involves a current flowing from island to island. In section 4.2 we assumed that the contributions to the conductance relaxation are additive, but a priori the macroscopic conductance (or resistance) is not a simple sum of individual conductances (or resistances). However, the effect of local resistance variations may be additive. The resistance of a macroscopic percolating system such as ours is that of the sub-network of critical resistances. These are all of the same order (of the order of $R_\square$) and a given variation of any of them induces the same small relative change of $R_\square$. We expect the variations of a number of them to have a cumulative effect on $R_\square$, at least as long as $\delta R/R$ remains small enough.

The mechanism of Cavicchi et al. can be applied to explain the formation of our $G(V_g)$ cusp. We assume that some of the critical resistances are isolated grains which, together with the gate electrode and the neighbouring leads of the percolation path, constitute single electron transistors (SETs). Taking a typical grain diameter of 50 Å and a tunnel barrier 10 Å thick we find that $\frac{e^2}{2c}$ is of the order of a few 100 K, so that grains can be subjected to Coulomb blockade at all our measurement temperatures. As long as the bias voltage is low enough, the discussion is not different from that of the electron boxes of Cavicchi et al. The dielectric polarization has the same effect on the SET islands potential distribution (or on the $V_D$ distribution, $V_D$ being the difference between the island and close plate potentials). The gaps appearing in the $V_D$ distribution near $+/-\frac{e}{2c}$ alter the macroscopic conductance in the same way as they alter the capacitance in the Cavicchi et al. configuration, and give rise to a dip in the $G(V_g)$ curves. At zero temperature the cusp width is determined by the amplitude of the dielectric relaxation around the islands. The latter can be non zero since it can occur by tunnelling. At $T = 0$ one thus expects a $G(V_g)$ cusp with a non zero "intrinsic" width.

Finite temperatures can modify the picture in several ways. For simplicity we restrict our discussion to the case of non activated polarisation processes [36]. The "soft" grains are now those islands with $V_D$ close to $+/-\frac{e}{2c}$ within $k_BT/e$. Their charge state thermally oscillates between two values, which results in a broadening of the $V_D$ distribution edges. This results in a broadening of the $V_D$ distribution gaps, and of the $G(V_g)$ cusp. In our experiments the cusp width is proportional to $k_BT$ in the whole temperature range, thus the thermal broadening is always larger than any intrinsic width. The measured $G(V_g)$ cusp width should be of the order of a few $\frac{c}{c_g^{grain}}k_BT$, where $c$ is the total capacitance of a typical SET island relative to its environment and $c_g^{grain}$ is its capacitance relative to the gate. Owing to the screening by the neighbouring grains, we estimate a typical $c_g^{grain}$ using the planar capacitor formula. For samples with a 1000 Å thick $Al_2O_3$ gate insulator we estimate $\frac{c}{c_g^{grain}} \approx 500$ and a cusp width of a few times 200 mV at $T = 4$ K, which is in agreement with the experimental value of 1 V.

In this "extrinsic" model, the electrons by themselves have no memory. Therefore the mechanism of the thermal memory (demonstrated in Figure 6 ), must be sought in the dielectric polarisation. A plausible origin may be as follows. At a finite $T_1$, the islands with $V_D$ close to $+/-\frac{e}{2c}$ within $k_BT_1/e$ oscillate between two charge states, their average charge depending on $V_D$. So does the electric field around them. The slow dielectric polarisation responds to the average electric field around the islands. Thus, islands which would be in the same charge state at $T = 0$, and would then be subjected to the same dielectric polarization, are subjected to different dielectric polarisations at $T_1$, which is determined by their $V_D$. When the sample is quenched from $T_1$ to $T_2 \ll T_1$, the islands charges essentially stop oscillating, but the distribution of dielectric polarisations is kept in its $T_1$ configuration, maintaining a broadened anomaly in the $V_D$ distribution, and hence in the $G(V_g)$ curves. This suggestion for the mechanism of the thermal memory should be further investigated and quantitatively checked.

The preceding discussion shows that the "extrinsic" model developed by Cavicchi et al. can be used to explain reasonably well the experimental findings we have reported. However a quite different interpretation can also be invoked, based on the Coulomb glass concept, which we now consider.

### 4.3.3 Electron Coulomb glass interpretation

The experiments on $InO_x$ have been interpreted in the quite different framework of the Coulomb glass [5, 37]. In brief, the Coulomb glass state results from the strong localisation of the carriers by disorder and the ill-screened Coulomb repulsions between them which strongly correlate their movements. The $G(V_g)$ cusp can be understood in the following way: the system has its minimum conductance in its ground state, so that an excitation to a higher state (like when changing the gate voltage) increases the conductance, and the relaxation to the new ground state at the new gate voltage is extremely slow. The main experimental finding supporting the idea that electron correlations are essential is the effect of doping: increasing the electron concentration in $InO_x$ widens the conductance cusp and significantly slows the relaxation, as qualitatively expected for the Coulomb glass [7].

The Coulomb glass ideas may also be applied to granular metals. The similarity between this system and continuous disordered insulators has been outlined [12]. The metallic grains play the role of localised states, the site energy disorder arises from the always present distribution of grains potential, and the Coulomb interaction between charged grains sets the interaction energy scale (of the order of a grain charging energy for nearest neighbours interaction).

On the theoretical side, the Coulomb glass and its experimental manifestations are now the subject of a renewed interest [38, 39, 40, 41, 42, 43]. Although the question of a finite glass transition temperature in real systems is still theoretically debated [41, 42, 44] and experimentally unanswered, there is a consensus that the relaxation of a Coulomb glass through the low lying metastable states involve multi-electron hops which can have arbitrarily long times.

In order to test the Coulomb glass interpretation, one needs theoretical predictions directly applicable to the experiments. Long time relaxations have been predicted to be logarithmic in accordance with experiments, but this is by no means a distinctive feature. The precise mechanism for the cusp formation is debated. The suggestion that it is a direct image of the formation of the Coulomb gap [9, 39, 45] can probably be ruled out. The relative conductance changes are rather small (at most 20%), while it should change by orders of magnitude upon the formation of the Coulomb gap. More over the cusp width is proportional to $k_BT$, unlike the Coulomb gap. Finally the latter is expected to essentially form in short times [38]). Pollak et al. [37] suggest that the perturbation caused by a change in $V_g$ mainly comes from a change in the localised sites energies, which destabilizes the system. On the other hand Lebanon et al. [46] suggest that the main effect on the conductance is due to the shift of the chemical potential within the Coulomb gap, induced by the injected carriers. The subsequent slow conductance relaxation (and the new cusp formation) corresponds to a rather small readjustment of the Coulomb gap to the new carrier concentration. These authors have made some predictions which can be tested. A simple criterion for the cusp width $\Delta V_g$ predicts:

$$e\Delta V_g = 2Ln(2)\frac{e^2 v_0 d}{C_g}k_B T - \frac{e^2 v_0 d E_c}{2C_g} \qquad (3)$$

where $d$ is the film thickness, $C_g$ the gate to film capacitance per surface unit, $v_0$ the bare density of states and $E_c$ the Coulomb gap width. A direct application to granular metals, with the disorder limited by the charging energy (hence $v_0 = \frac{n_g}{E_c}$ where $n_g$ is the grain surface density and $E_c$ is now the charging energy) gives:

$$e\Delta V_g = 4Ln(2)\frac{c}{c_g^{grain}}[k_B T - k_B T_0] \qquad (4)$$

with $k_B T_0 = \frac{E_c}{4Ln(2)}$ ($c_g^{grain}$ is the planar capacitance between the gate and a grain like previously). The slope of the linear dependence of the cusp width with temperature is thus consistent with our measurements, and similar to that expected from the Cavicchi-Silsbee model. The negative constant term given by the prediction is not observed in our experiments, but recent numerical simulations [47] show that the cusp width may tend to a small finite value as $T \rightarrow 0$, which can be consistent with our experimental data.

**4.3.4 Open questions**

The origin of our cusp and of its slow dynamics is still an open question. Our experimental results are very similar to the ones reported in InO$_x$. While the "extrinsic" interpretation seems more easily applicable to granular metals and the Coulomb glass model to InO$_x$, it seems to us that both explanations could be applied to both systems. In the case of InO$_x$, the effect of the electron doping level on the cusp and its dynamics favours the electron Coulomb glass. However we should keep in mind that the doping is obtained by increasing the oxygen deficiency, and the nature and number of "extrinsic" ionic degrees of freedom (TLS etc…) most probably depend on the quite high vacancy concentrations.

We also remark the striking point that the slow relaxation is observed for samples relatively close to the metal to insulator transition. For instance it was observed in a sample with $R_\square(4K) \approx 100$ k$\Omega$ and $R_{4K}/R_{300K} =$ 8,6. In these, the effects are more difficult to measure because the *relative* conductance changes become tiny (in the case mentioned the "steady" cusp amplitude was 0.03 %), but the relaxation times seem to be as long as in more insulating samples. This behaviour is expected in the "extrinsic" context. It is more surprising in the electron Coulomb glass framework, but it may arise from the intrinsic inhomogeneity of the systems. One can perhaps imagine percolating conducting channels, influenced by the slow relaxation of nearby electronic glassy zones, so that the electron glass may manifest itself in macroscopically metallic samples. Bielejec et al. [48] reported on the observation of a slow conductance relaxation of *metallic* ultra-thin quenched condensed Al films, following a cool-down at He dilution temperatures. It is difficult to ascertain whether the phenomena occurring in that case are the same as the ones we studied, but clearly studies at lower temperatures deserve further experimental attention.

The distinction between the extrinsic ionic scenario and the Coulomb glass is all the more difficult that one can expect qualitatively similar effects from both. For instance studies of slow relaxation phenomena in structural glasses have shown that the TLS are rather complex, their transitions involving the rearrangement of several ions/atoms. They may display quite complex dynamics, and their rather low activation energies are in the same range as the ones of electron correlation or Coulomb gap energies. Thus, from an experimental point of view, we feel that the question remains open, whether we have the simple situation of a collection of independent extrinsic degrees of freedom influencing the current carrying electrons (giving rise to a situation similar to the one suggested recently for another system [49]), or if we observe the result of the highly correlated dynamics of the electron Coulomb glass. Up to now we see no clear-cut experimental result which designates the proper interpretation without ambiguity.

The study of small samples may be a route for a better understanding of the physics involved. A more detailed account of the mesoscopic effects observed in micrometric samples will be given elsewhere, we just want to point here to a rather curious observation.

When the samples dimensions are small enough (typically below tens of microns), the field effect curves $G(V_g)$ display reproducible fluctuations, like the ones shown in Figure 13. At liquid He temperature, they are very stable, their pattern can only be changed by heating the samples. As shown in Figure 14 the $G(V_g)$ cusp is superposed on the reproducible fluctuations. When the cusp is changed (for instance like in Figure 14 or in a TDE), the fluctuations are essentially unchanged. Whatever the origin of the reproducible fluctuations is (Coulomb blockade, geometrical or quantum effects), they should be sensitive to changes of the local electric fields and should thus be good sensors of the "extrinsic" dynamics. Their stability at low T could then indicate that the "extrinsic" degrees of freedom are frozen, and that the cusp and its dynamics do not depend on these (thus favouring the Coulomb glass scenario).

However, since the cusp relaxation can be induced by electric fields, it most probably involves charge and electric field redistributions. It is then quite surprising that the reproducible fluctuations are not sensitive to these redistributions. The fact that the cusp and the fluctuations are simply superposed and seem to have independent dynamics is thus quite mysterious to us. The resolution of that paradox would probably help to understand the origin of the $G(V_g)$ cusp and of its glassy behaviour.

## 5. Summary

Electrical conductivity measurements have shown that granular Al thin films do not attain equilibrium in laboratory times after they have been cooled from room to liquid helium temperature. The slow relaxation towards equilibrium is manifested by the evolution of a field effect anomaly, which displays a thermal and applied gate voltage memory. The relaxations observed are logarithmic in time, and display simple aging in $V_g$ induced relaxations. The relaxation phenomenology (simple aging and master curves) is the same as that of a collection of independent "degrees of freedom" whose relaxation has a cumulative effect on the sample's conductance, and occurs according to a log-flat distribution of characteristic times. The absence of temperature dependence of the dynamics suggests that it proceeds by quantum tunnelling. The phenomenological description of the relaxation curves does not say much about the physics involved. It appears that the experiments can be interpreted in terms of an "extrinsic" scenario involving the influence of slow ionic movements on the electronic transport, or in terms of an intrinsic glassiness of the electrons.

In order to progress on these questions, one needs more theoretical predictions which can be experimentally tested. In particular, almost all theoretical studies of the electron Coulomb glass are performed in the classical (highly localized) limit, while experiments are performed in the quantum limit. On the experimental side, insights in the physics may be gained from a systematic search of similar out of equilibrium phenomena in a wider variety of systems, in order to identify the conditions under which they are present. The study of smaller samples, when mesoscopic effects become important, can also give some insight in the physics at work. Such studies are presently under way, and will be reported later.

**Acknowledgements**

We wish to acknowledge numerous very fruitful discussions with Z. Ovadyahu, M. Müller and M.V. Feigel'man, as well as with L. Ioffe, K. K. Likharev, M. Ortuno, M. Pollak and B. I. Schklovskii.

**Appendix A**

In this appendix we recall the basic mechanism proposed by Cavicchi et al. [25], based on a first suggestion by Lambe et al [24], to explain the formation of a cusp in capacitance measurements. The authors discuss the differential capacitance of a plane capacitor which contains small isolated metallic islands inside its insulator, close enough to one of its two plates so that electrons can tunnel back and forth between that plate and the islands (Figure. 1.A). In today's terminology the islands constitute a parallel array of "electron boxes", each of which adds its own contribution to the differential capacitance $C$ measured as a function of an applied DC voltage $V_s$.

As is well known, when $k_B T \ll \frac{e^2}{2c}$ and the junction resistances are high enough ($c$ is the island capacitance defined in Figure 1.A) Coulomb blockade dominates the charge transfer to the islands. Their charge states are quenched, except when $V_s \approx e/2c_I \pm ne/c_I$ ($c_I$ is defined in Figure 1.A). When this condition is met, they can fluctuate between two degenerate charge states and contribute to the differential capacitance. Thus their contribution to $C(V_s)$ as a function of the bias is a periodic series of peaks. Cavicchi et al. recall that in fact the periodic contributions of the various islands are not in phase. One reason is that the islands work functions can differ markedly from one another. Moreover they may be influenced by various stray charges. The result is that their Fermi levels (or potentials) are broadly distributed, and their periodic contributions to $C(V_s)$ are correspondingly shifted. The sum of these uniformly dephased periodic contributions is a flat function, thus the $C(V_s)$ curve should show no cusp. Note that in fact, since the Fermi energy difference between an island and the nearby electrode ($eV_D$) cannot exceed $e^2/2c$, one has a rectangular distribution $N(V_D)$ delimited by $+/-e/2c$. If for some reason one island had $|V_D|>e/2c$, charge transfers would occur, which would shift $V_D$ by increments of $e/c$ until $|V_D|<e/2c$ (steps 1→2 in Figure 3.A) Anyhow the rectangular $N(V_D)$ still gives a flat $C(V_s)$. The periodic contributions of islands with different $V_D$ are sketched in Figure 2.A.

But these expectations are contradicted by the experiments wich reveal a cusp in the $C(V_s)$ curves (centred on $V_s = 0$ if the system was cooled in that condition). To explain it, one must find a mechanism which modifies the rectangular $N(V_D)$ distribution, and partially rephases the individual periodic contributions to $C(V_s)$. One way to achieve this is to depress $N(V_D)$ around its edges $V_D = +/-e/2c$, since the islands with these $V_D$'s contribute to $C(V_s)$ around $V_s = 0$ (see Figure 2.A) [50]. Actually, were all the islands identical, a periodic series of dips would result in $C(V_s)$ at $V_s = \pm ne/c_I$, but owing to the broad distribution of $c_I$, these are damped and there only remains a dip at $V_s = 0$, as is generally observed in experiments.

If $V_s$ is set to another value, the $V_D$ of the islands are shifted, the depressions in the $N(V_D)$ distribution are not around $+/- \frac{e}{2c}$ anymore, so that they are erased and new ones are created at the proper location. This is what happens in TDEs.

Cavicchi et al. propose a mechanism for the depletion of $N(V_D)$ at $V_D = +/-e/2c$. The slow modification of $N(V_D)$ is caused by an interplay between a slow polarisation of the dielectric material surrounding the islands and the discrete changes of island charge states it induces. The mechanism is illustrated in Figure 3.A, where we schematize step by step how $N(V_D)$ is altered. First in situation (1) we recall what the situation would be if all the islands were neutral: their $V_D$ would be broadly distributed. In (2), those islands which had $|V_D|>e/2c$ in (1) have lost or gained electrons so that they now all obey $|V_D|<e/2c$, giving the rectangular distribution. Most of the islands are now charged. The charging of the islands has incremented the electric fields around them, to which the dielectric responds by polarizing. From situation (3) on, we isolate those islands which charge is $Q = e$ in situation (2). The dielectric polarization shift their $V_D$'s upward as shown in (4). Those islands which $V_D$ has been shifted above $e/2c$ loose one more electron (then $Q = 2e$) so that they enter [$-e/2c$, $+ e/2c$] again, filling the gap near $V_D = -e/2c$ (5). The increase of electric field created by the extra charging again polarizes the dielectric surrounding the islands with $Q = 2e$, which potential is once more shifted upward (6). The result of that series of processes is a gap near $V_D = -e/2c$ in $N(V_D)$.

It is easily seen that all positively charged islands contribute to the same gap at $-e/2c$, while the negatively charged islands contribute to a symmetric gap at $+e/2c$. The width of the gap is determined by the amplitude of the dielectric polarisation induced by an increment of electric field. According to that model, the existence of the $C(V_s)$ anomaly and its long time evolution are thus not inherent to the electron dynamics in the system (jumps to / from the islands), but to the slow dynamics of the polarisation processes of the dielectric. The precise nature of these polarization processes is generally not known.

**Figure captions**

**Fig. 1:** XRD spectra of the (111) and (200) lines of Al measured on a series of four 5000 Å thick samples, evaporated at a rate of 1.8 Å/sec, in different oxygen pressures: (a): $P_{O2} \approx 10^{-6}$ mbar; (b): $P_{O2} \approx 1.2 \; 10^{-5}$ mbar; (c): $P_{O2} \approx 2 \; 10^{-5}$ mbar; (d): $P_{O2} \approx 3 \; 10^{-5}$ mbar. To give an idea of the electrical properties of the films, the $R(4K)/R(300K)$ values are: (a): 0.28 ; (b): 0.93 ; (c): 1.53 ; (d): $7.7 \; 10^{7}$.

**Fig. 2:** Normalised resistance versus temperature curves for a series of 200 Å thick granular Al samples. In the inset is shown the conductance versus bias voltage curve of the $R(4K)/R(150K) \approx 100$ sample.

**Fig. 3:** Typical field effect curve $G(V_g)$ obtained at 4 K with a MOS device whose granular Al channel has $R(4K)/R(300K) = 2 \; 10^3$ and $R_\square(4K)= 32$ MΩ.

**Fig. 4:** Effect of the temperature on the conductance "cusp". The FWHM and amplitude at each temperature are extracted from a lorentzian fit (a), which gives a good description of the cusp (except at the lowest temperatures). The FWHM increases linearly with $T$ (b), while the amplitude vanishes (like $1/T^2$ in the case of this sample) (c). Sample with $R_\square(4K)= 2$ GΩ)

**Fig. 5:** Cusp relative amplitudes (in %) measured for three different samples, of varying insulating strength. From left to right: $R_\square(4\;K) = 200$ kΩ, $R_\square(4\;K) = 300$ MΩ and $R_\square(25\;K) = 1$ TΩ.

**Fig. 6:** Slow relaxation of the $G(V_g)$ cusp after cooling a sample. (a): evolution of the cusp shape after cooling from high temperature to 4 K. From top to bottom: curves taken 19 min, 38 min, 76 min and 380 min after the cooling (the curves are vertically shifted for convenience). (b): logarithmic time evolution of the cusp depth after the cooling. Same sample as in Figure 4.

**Fig. 7:** "Two dips experiment": the sample is first equilibrated with $V_g = 0$ (step one, top curve); $V_g$ is set to 2.5 V for 3 hours (step 2), which results in a second dip (second top curve). After $V_g$ is set back to zero, the second "dip" vanishes (step 3, five lower curves, measured each 24 minutes). The curves are shifted for clarity ($T = 4$ K). Sample with $R_\square(4K)= 1,3$ GΩ.

**Fig. 8: (a)** time evolution of the first (increasing) and second (decreasing) cusp amplitudes in step 3 of the TDE. The sum of the two cusp amplitudes remains constant. **(b)** illustration of simple aging and of the master curve in step 3 of the TDE. Shown is the decreasing amplitude of the second cusp which was written in step 2 during (from left to right): 1080 sec, 5400 sec, 27000 sec. The inset shows the perfect collapse of the curves once plotted as a function of the reduced time $t/t_w$ ($T = 5$ K). Same sample as in Figure 4.

**Fig. 9 :** time evolutions of the first and second cusp amplitudes during step 2 of a TDE, for two durations of step 1 (300 s and 3000 s). Decreasing curves: amplitude of the first cusp being erased (left curve corresponds to shorter $t_w$); increasing curves: amplitudes of the second cusps being written. Sample with $R_\square(4K)= 500$ MΩ.

**Fig. 10:** time evolution of the second cusp amplitude in step 3 of the TDE, performed at different temperatures (in order of decreasing amplitudes: 4.5 K, 5.5 K, 6.5 K, 8 K, 12.5 K, 15 K and 20 K). The inset shows that the vertically rescaled curves all collapse on the same master curve whatever the temperature is. Same sample as in Figure 4.

**Fig. 11:** Time evolutions of the second cusp amplitude in step 3 (erasure) of the TDE, performed at the same and at a higher temperature than step 2 (writing). The writing was performed at $T = 5$ K, and the erasure was performed at 5 K in one case, at 8 K in the other case. Note that the second curve is vertically scaled to the first one. We see that the erasure process is not accelerated when performed at higher temperature than the writing. Same sample as in Figure 4.

**Fig. 12:** time evolution of the second cusp amplitude in step 3 of the TDE, as measured (diamonds) and calculated as in the text in the continuous $\tau_i$ distribution limit (line).

**Fig. 13:** reproducible conductance fluctuations of a small sample (0.1 x 0.5 μm²), the time interval between each curve shown is 6h40min (curves are shifted for clarity).

**Fig. 14:** Formation of a cusp at $V_g = 0$, superposed to stable conductance fluctuations in a small sample (≈30 x 30 μm²). Prior to $t = 0$, the cusp was "eliminated" by imposing a randomly variable gate voltage (curves are shifted for clarity).

**Fig. 1.A** "Tunnel" capacitor, studied by Lambe et al. [24] and Cavicchi et al. [25], and its "electron box" equivalent: $c_R$ is a typical tunnel junction capacitance (island to close plate capacitor), $c_I$ is the typical island to distant plate capacitance, we define $c = c_R+c_I$.

**Fig. 2.A** Schematized contributions to the differential capacitance $C$ of islands of different $V_D$. Islands of a given $V_D$ contribute a periodic series of sharp peaks (low $T$ limit). The phase of this series of peaks is determined by $V_D$. The examples shown are for $V_D = 0$ (grey peaks) and $V_D = +/- e/2c$ ("soft" islands, black peaks).

**Fig. 3.**A Mechanism for the formation of a gap at the edges of the rectangular $N(V_D)$ distribution. In situation (1) we show a broad island potential distribution, prior to charge transfers with the close plate. The rectangular distribution shown in situation (2) is obtained once islands previously having $|V_D|>e/2c$ have been charged until their $|V_D|<e/2c$. For simplicity, we then restrict the discussion to islands having a charge $Q=e$ in (2) (their $N(V_D)$ is isolated in (3)). The electric field created by the transfer of the charge $Q=e$ polarizes the dielectric, which shifts the island potentials upwards (situation (4)). The islands which have once more $|V_D|>e/2c$ lose one more electron ($Q=2e$), which shifts their $V_D$ by $-e/c$, thus filling the gap near $V_D=-e/2c$ (5). The extra electric field created by this second charge transfer increases the dielectric polarization, shifting again the $V_D$ of these islands upward, finally creating a gap in the $V_D$ distribution near $V_D=-e/2c$ (6). It can easily be seen that all islands with a positive charge in (2) contribute to the same gap in $N(V_D)$, while these with a negative charge create a symmetrical gap near $V_D=+e/2c$.

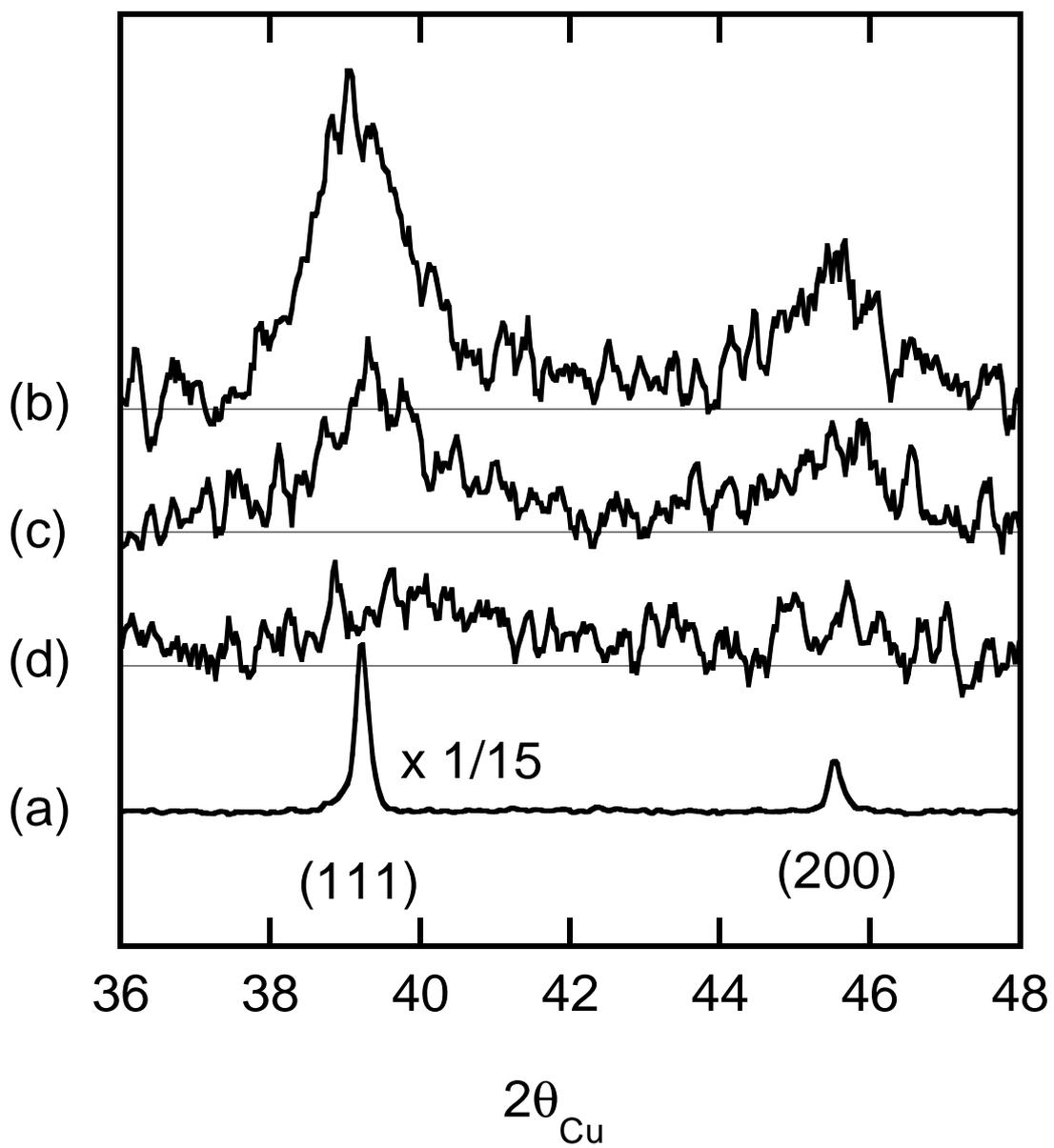

**Figure 1**

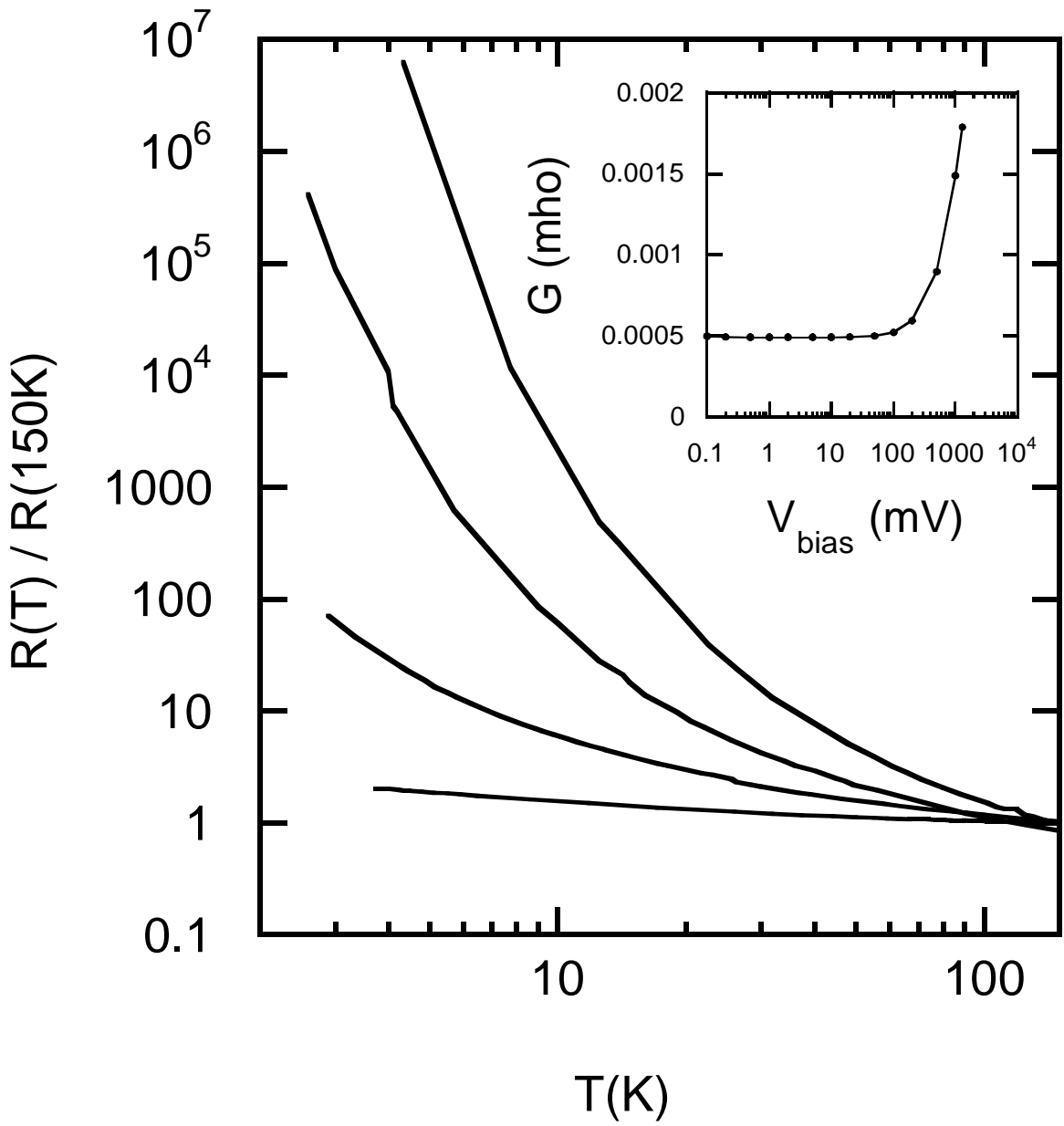

**Figure 2**

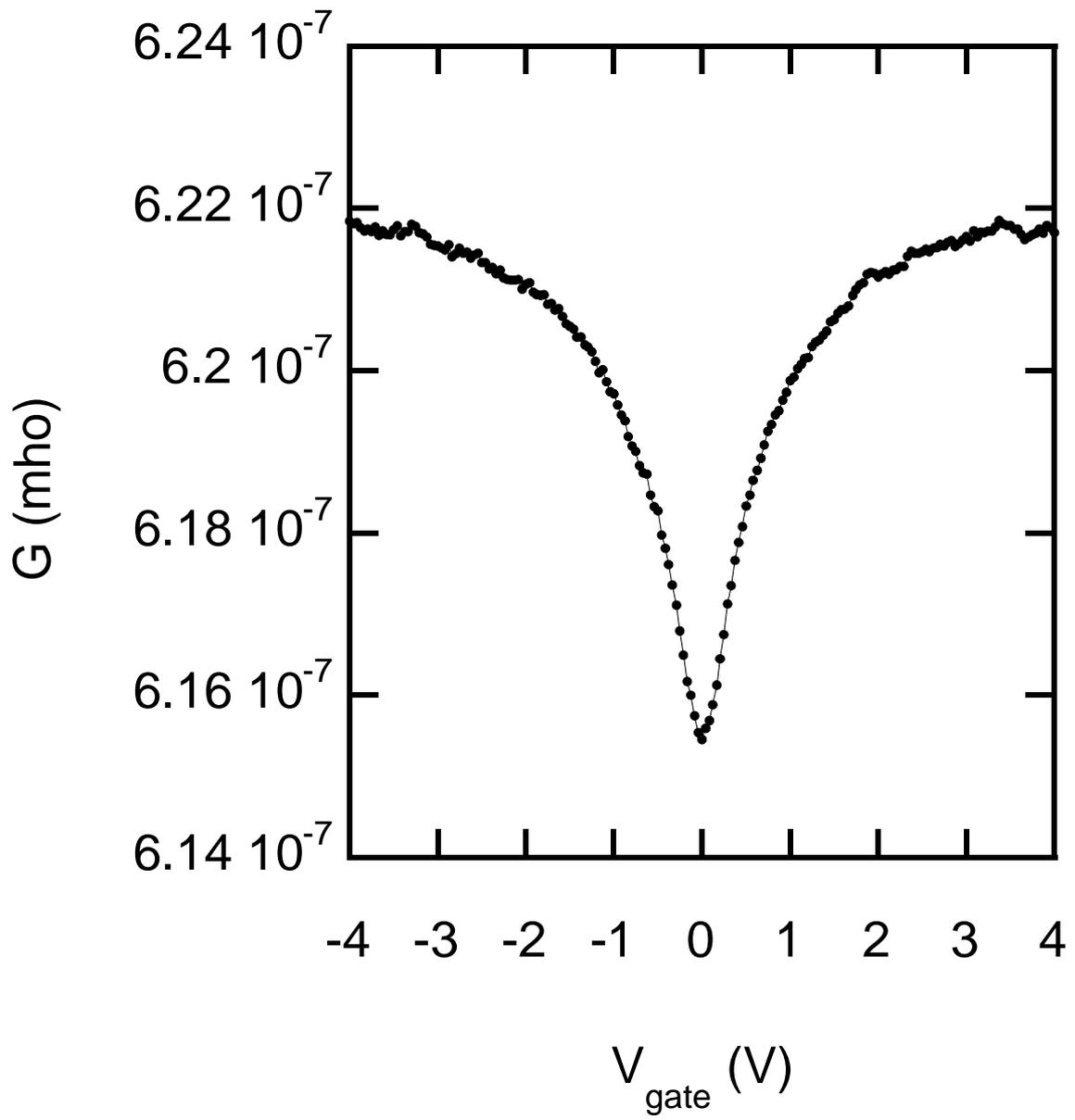

**Figure 3**

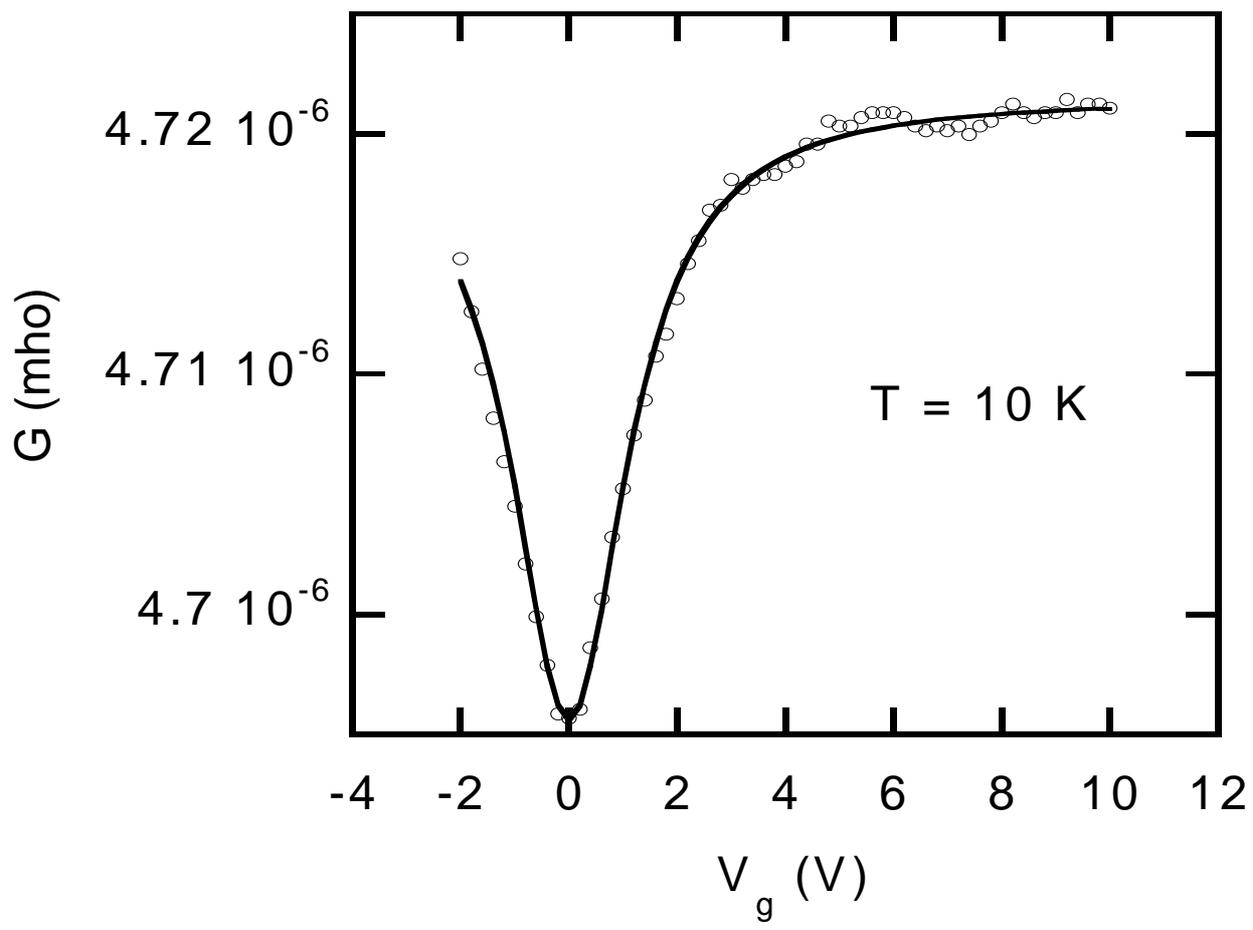

**Figure 4-a**

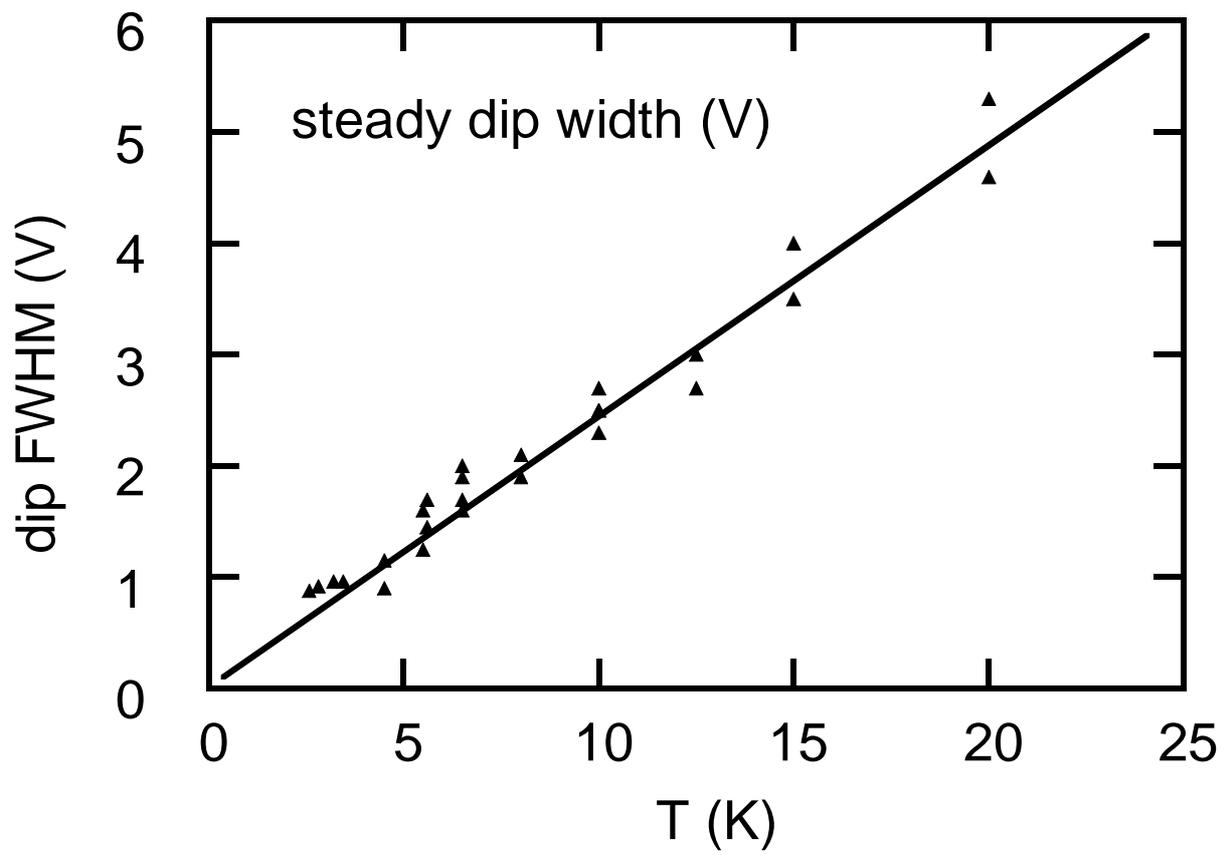

**Figure 4-b**

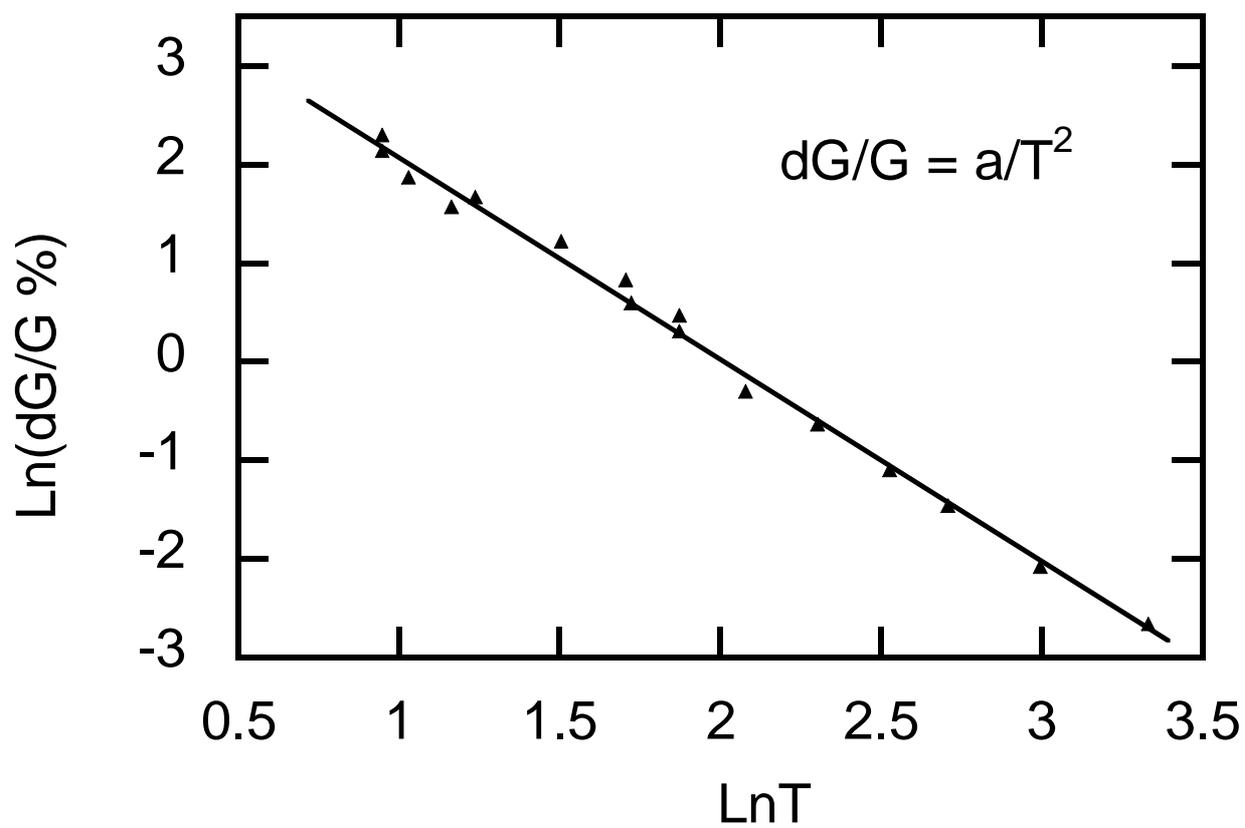

**Figure 4-c**

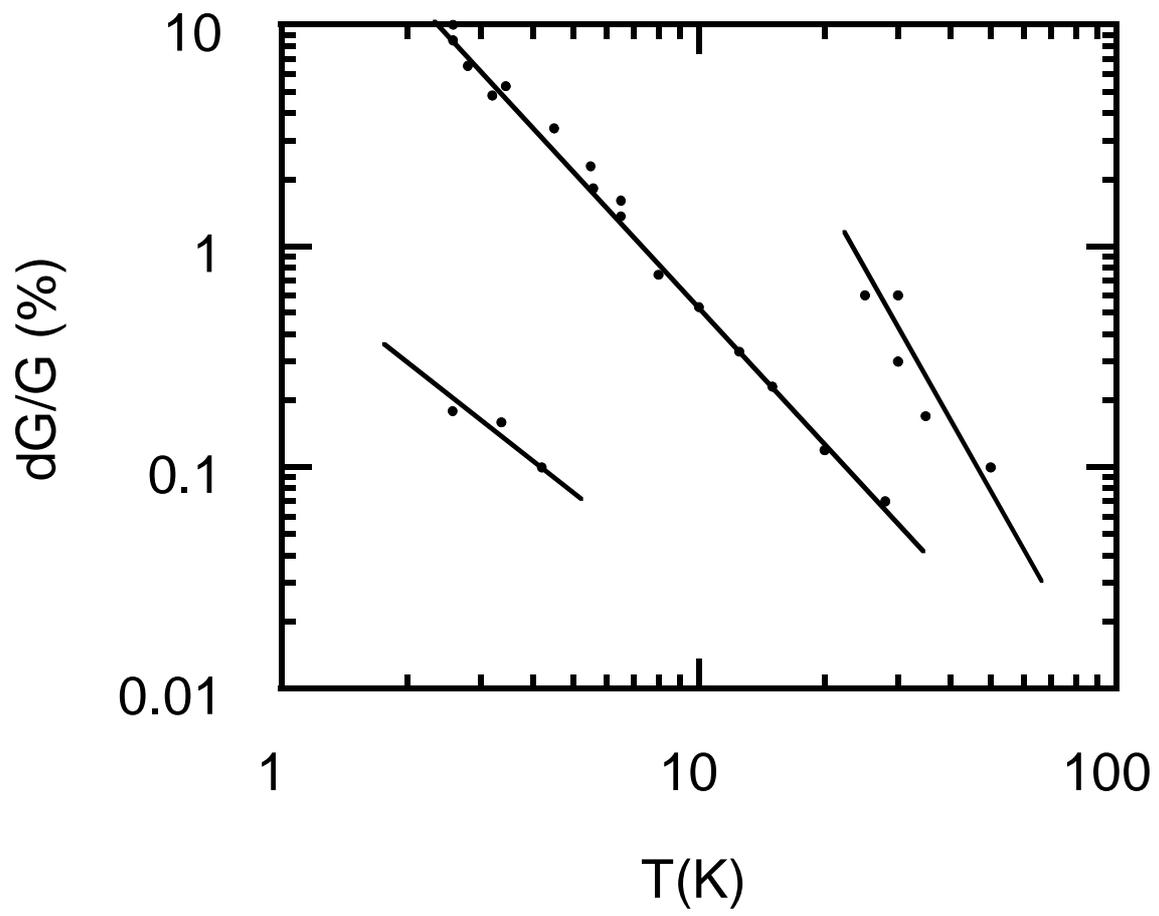

**Figure 5**

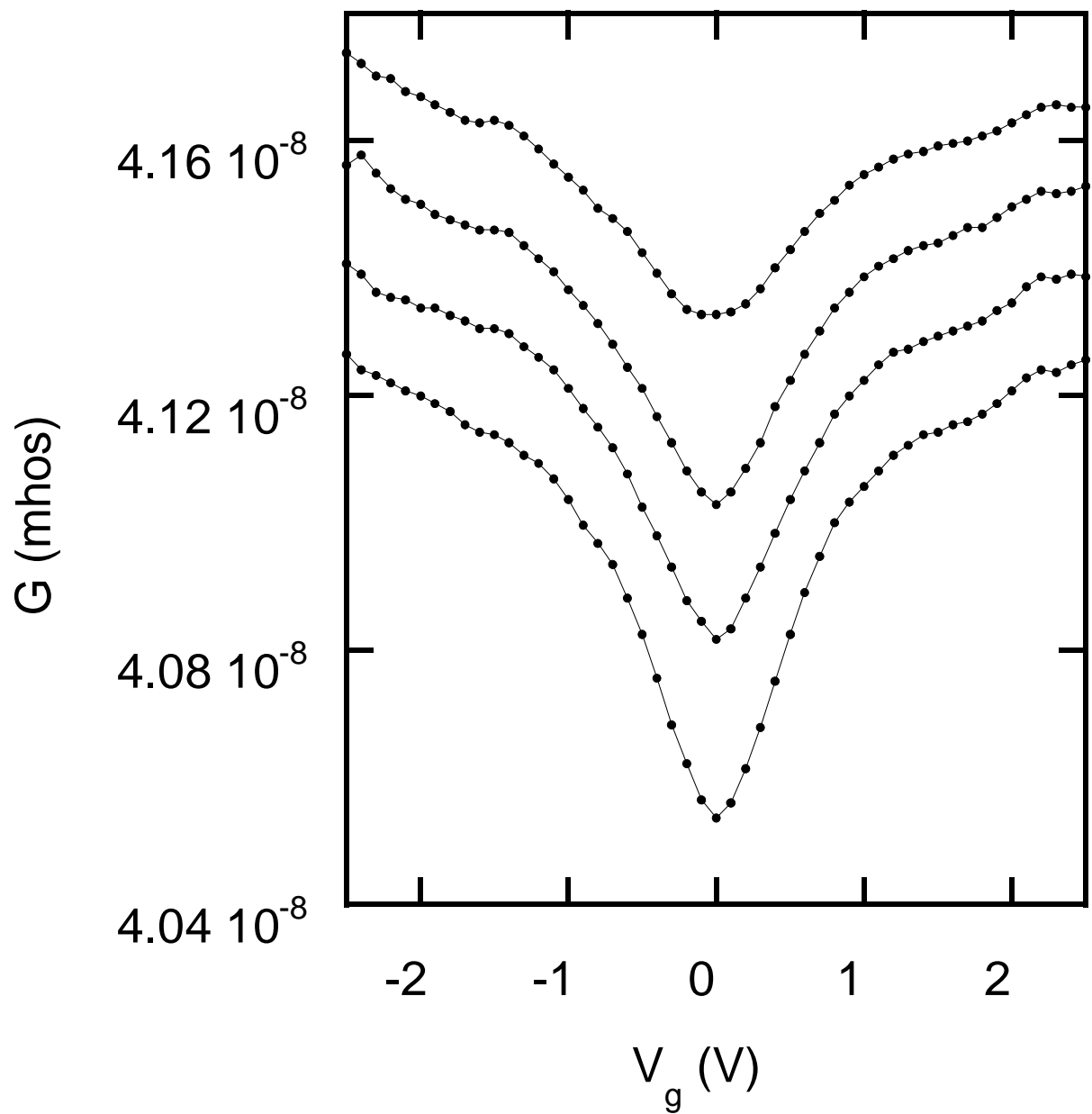

**Figure 6-a**

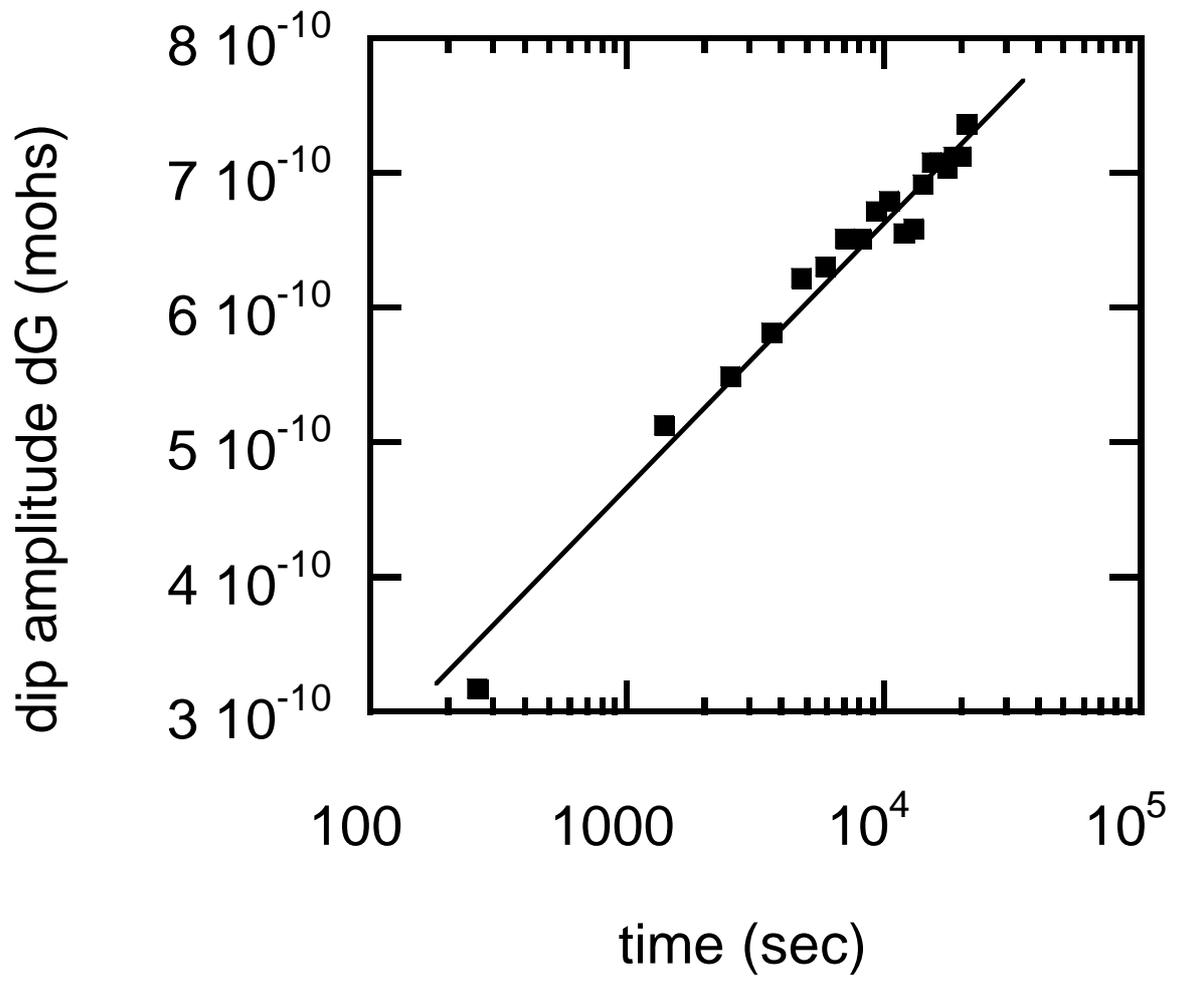

**Figure 6-b**

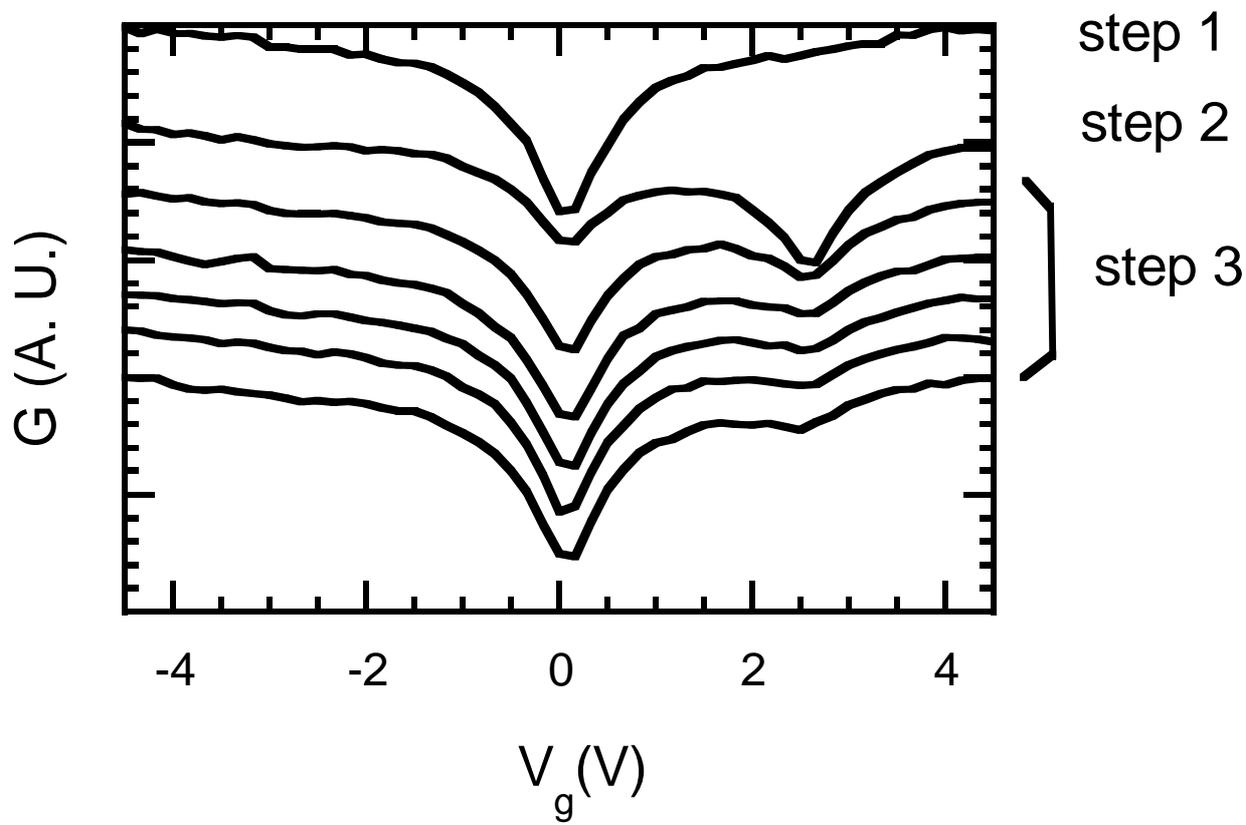

**Figure 7**

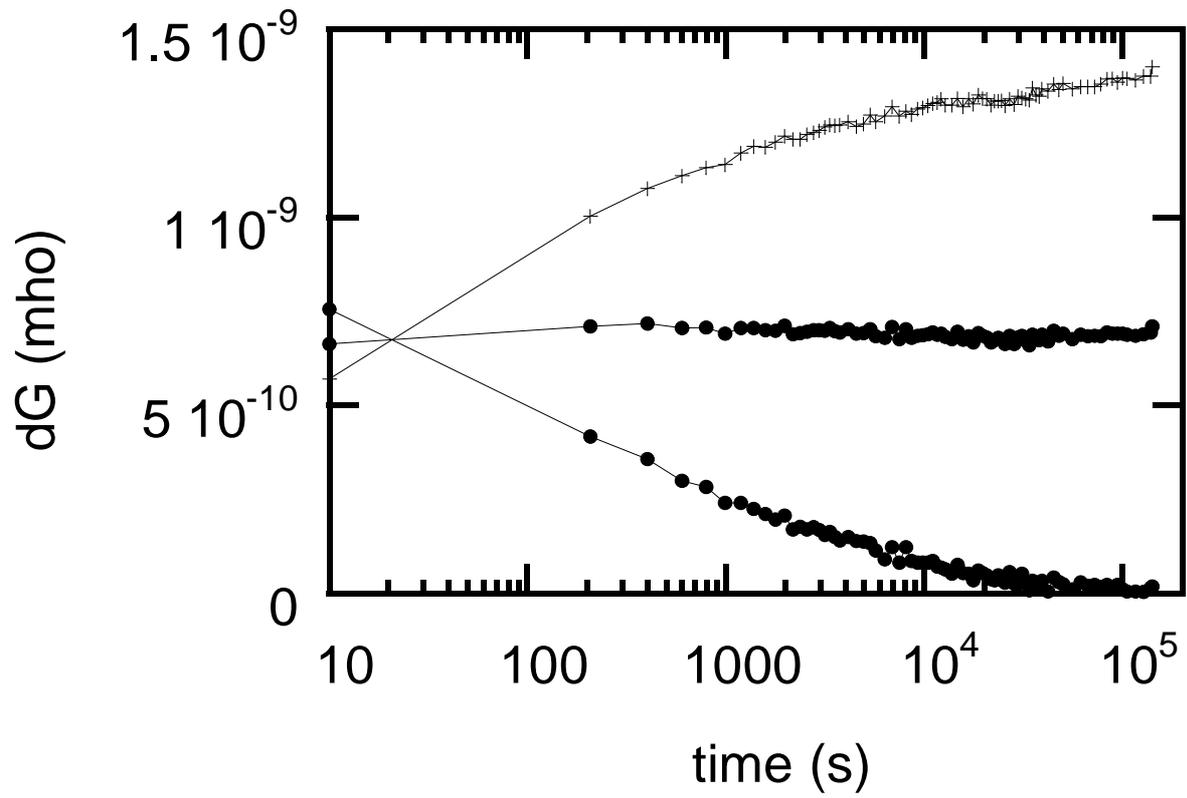

**Figure 8-a**

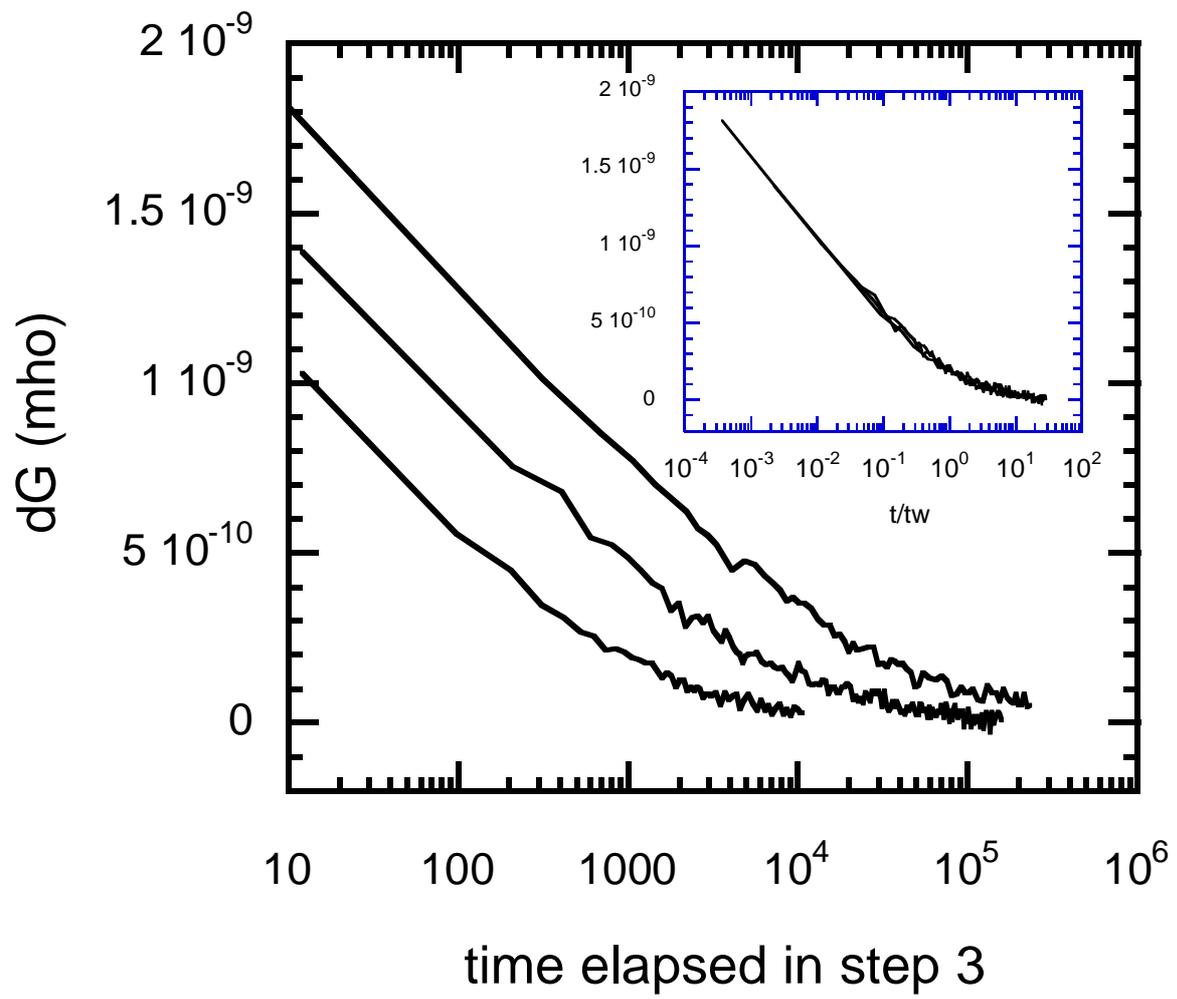

**Figure 8-b**

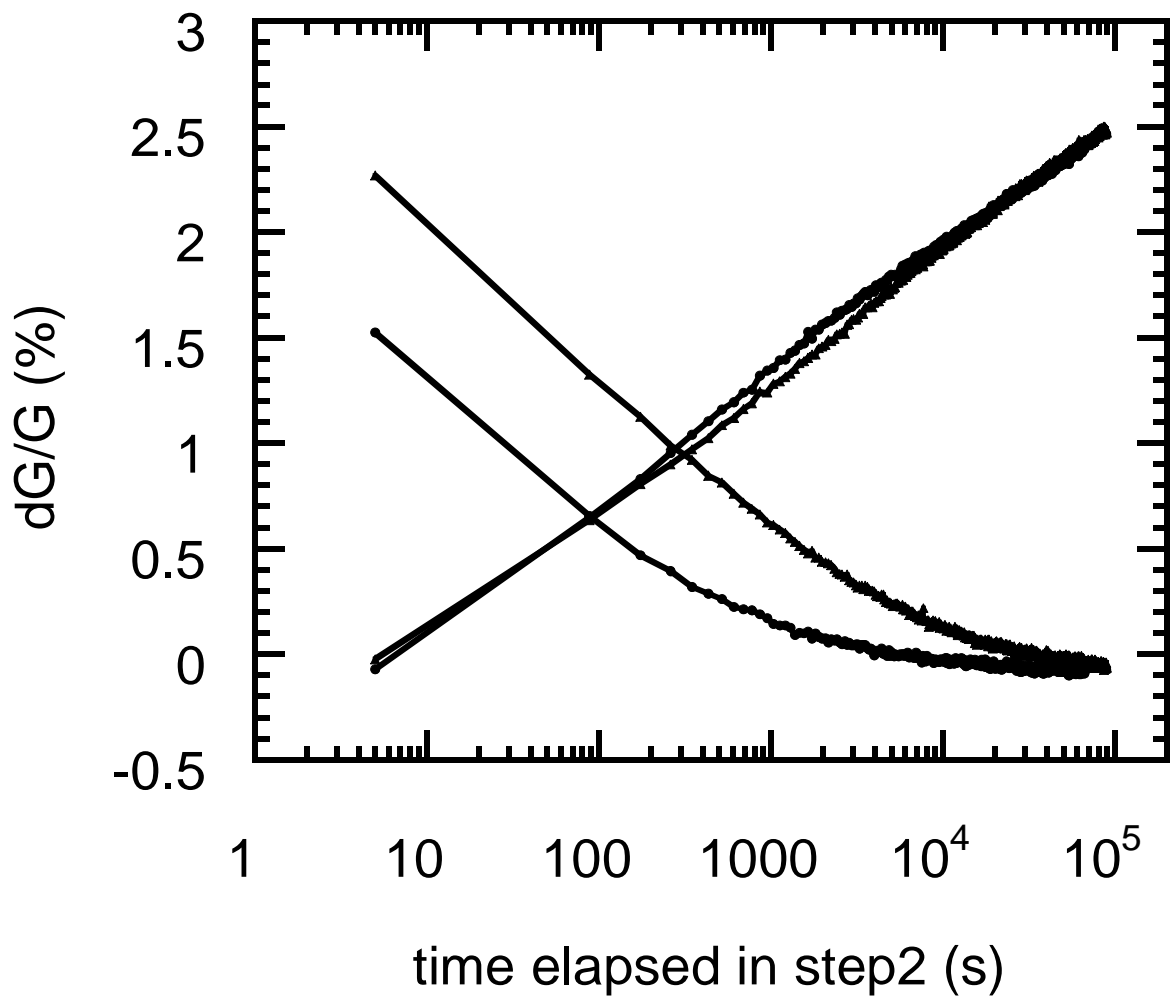

**Figure 9**

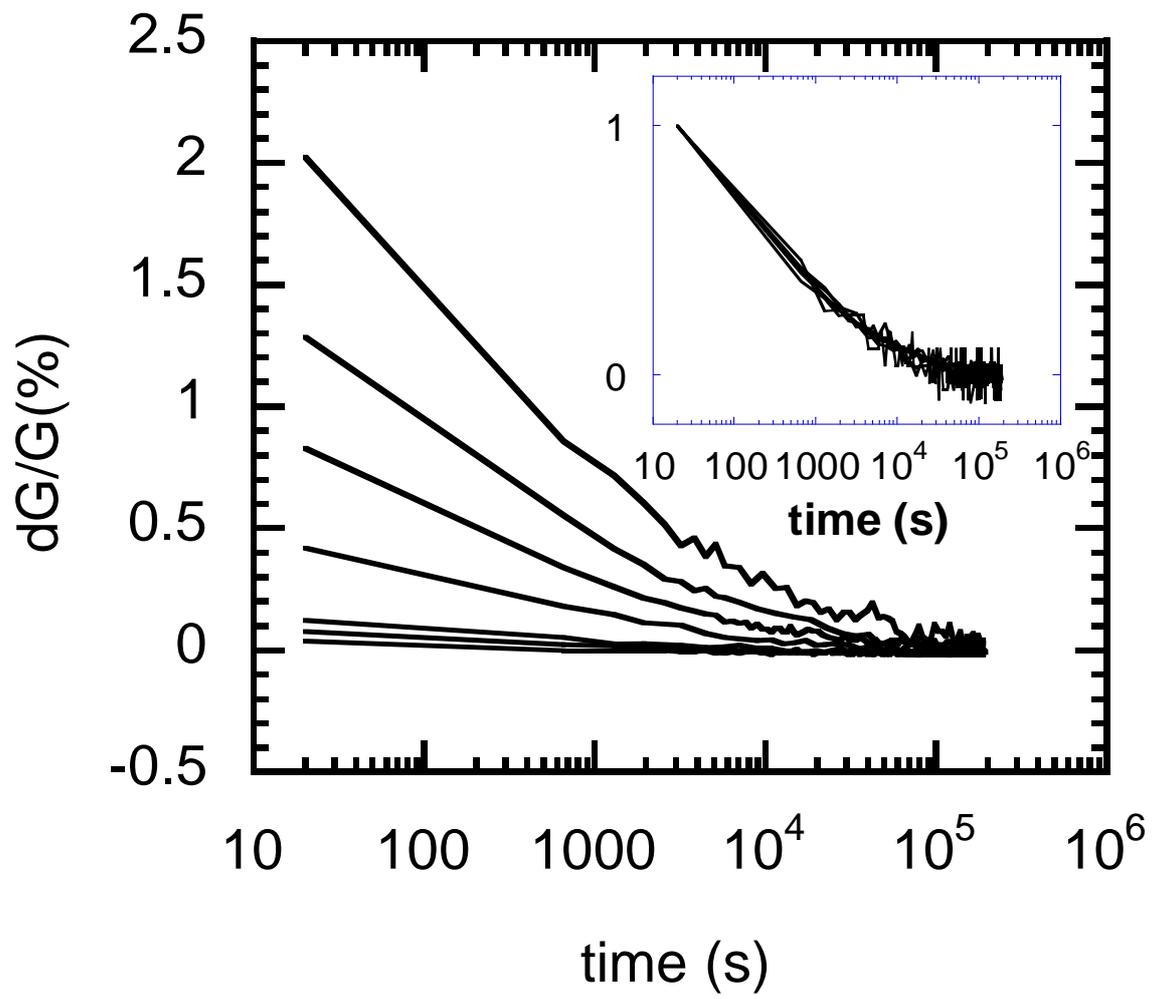

Figure 10

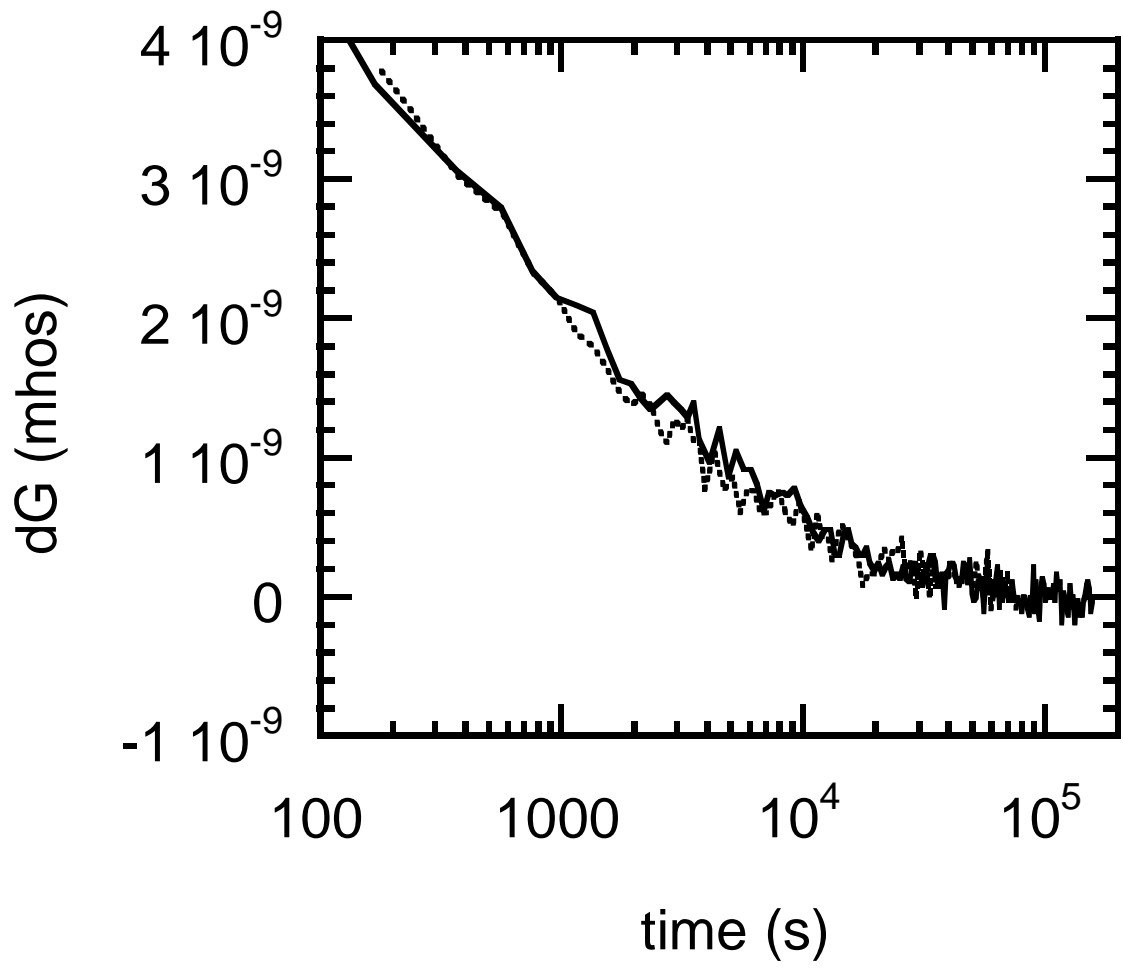

**Figure 11**

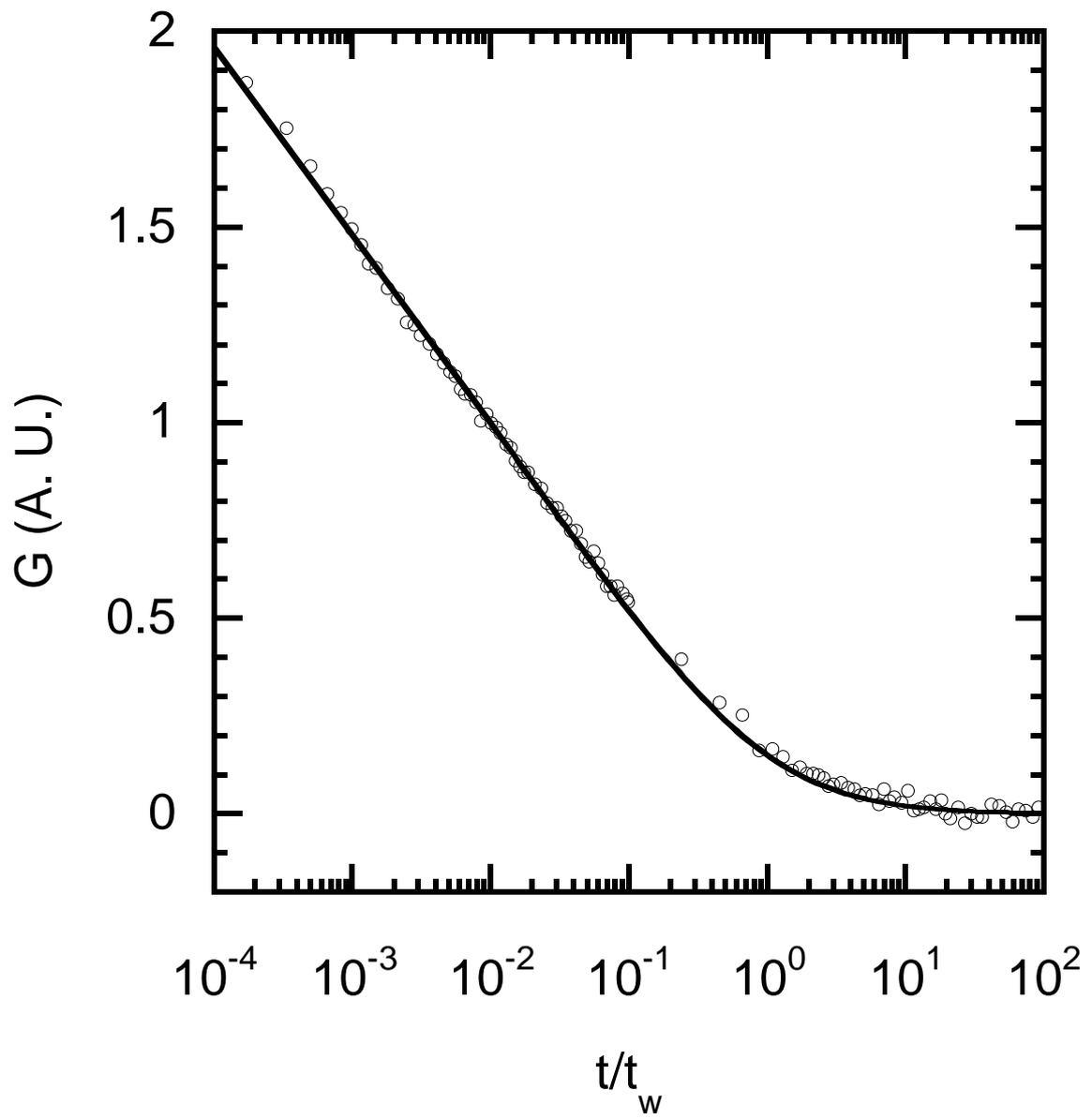

**Figure 12**

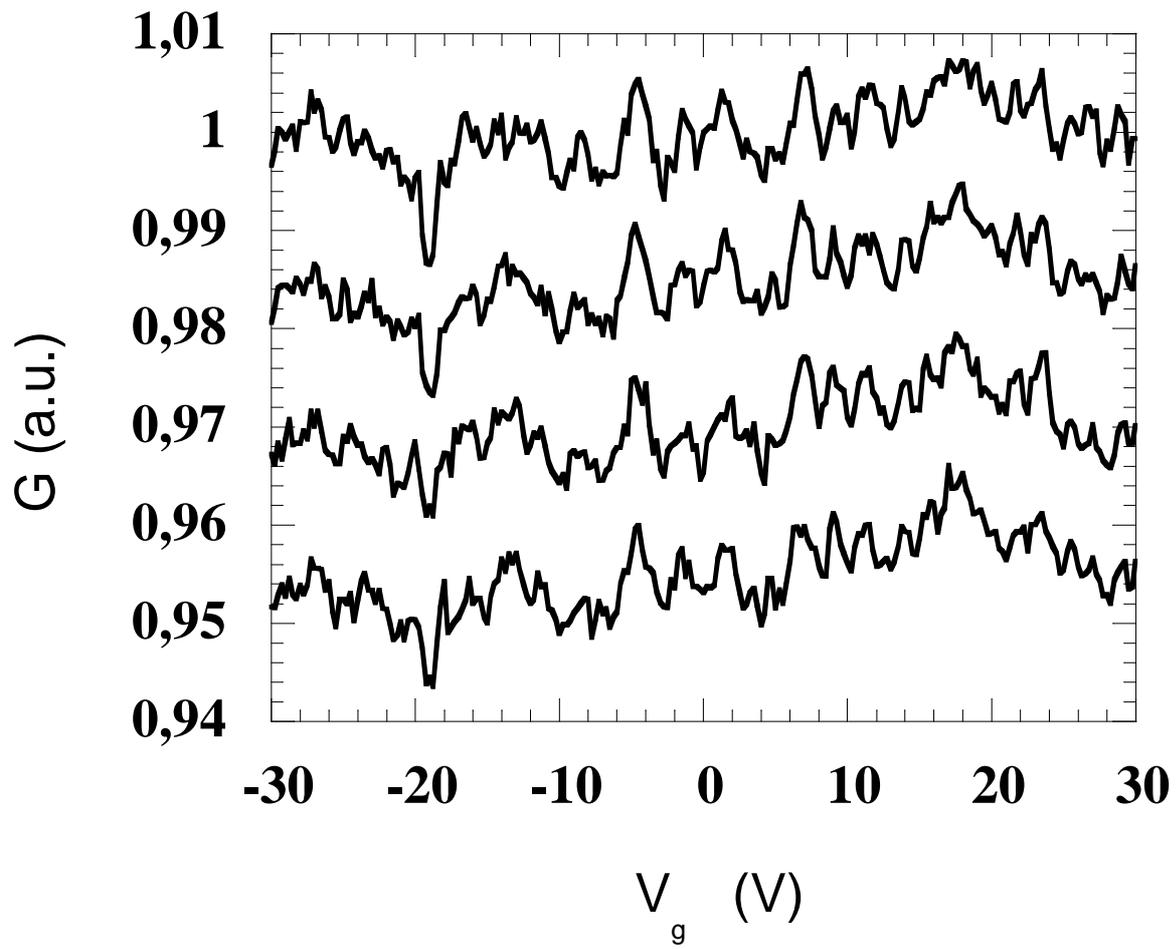

**Figure 13**

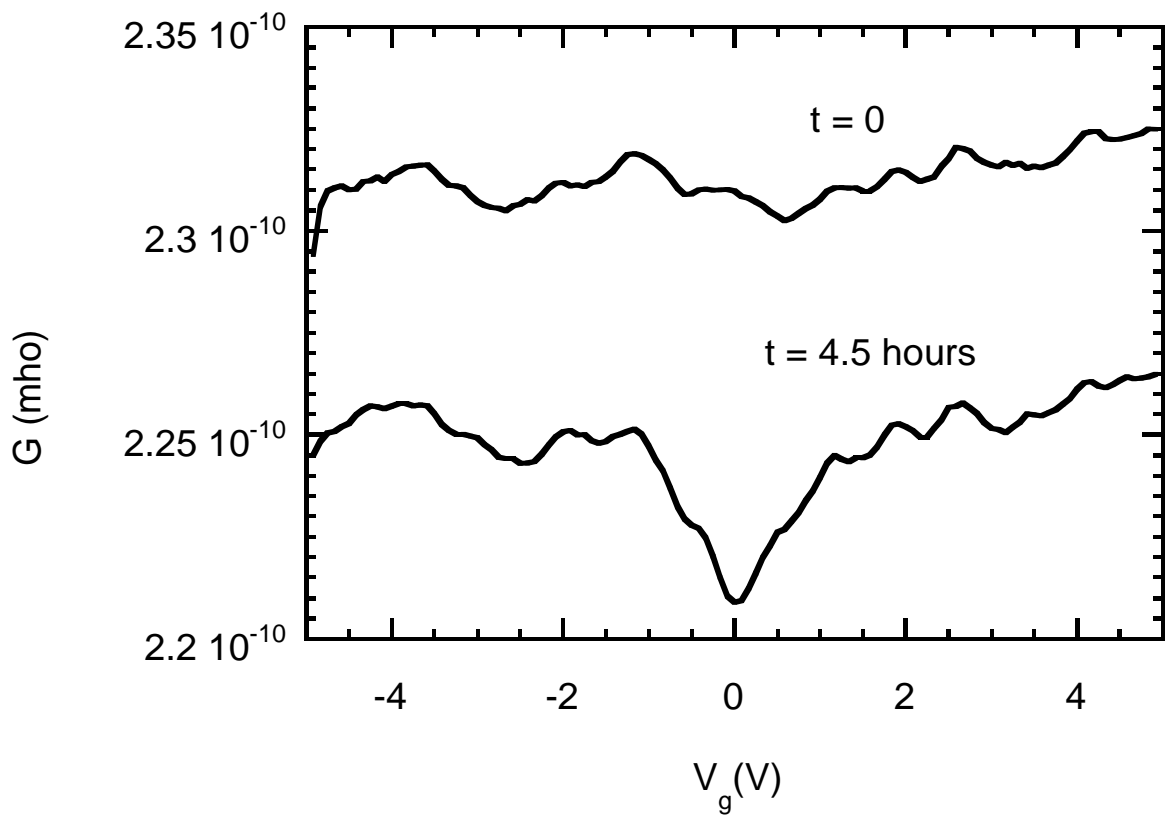

**Figure 14**

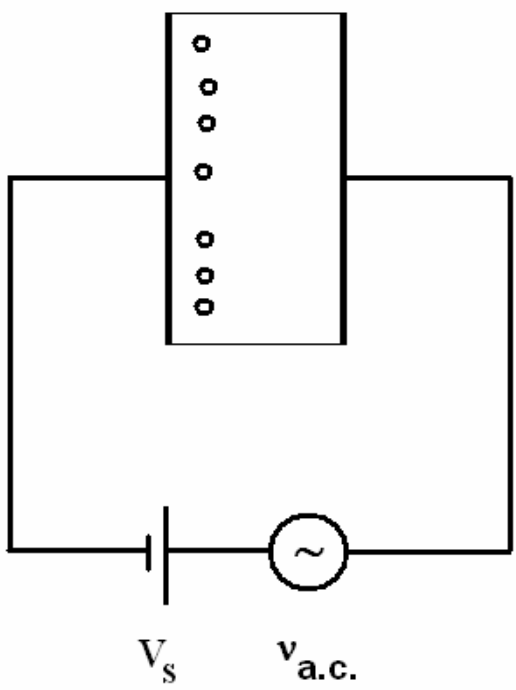 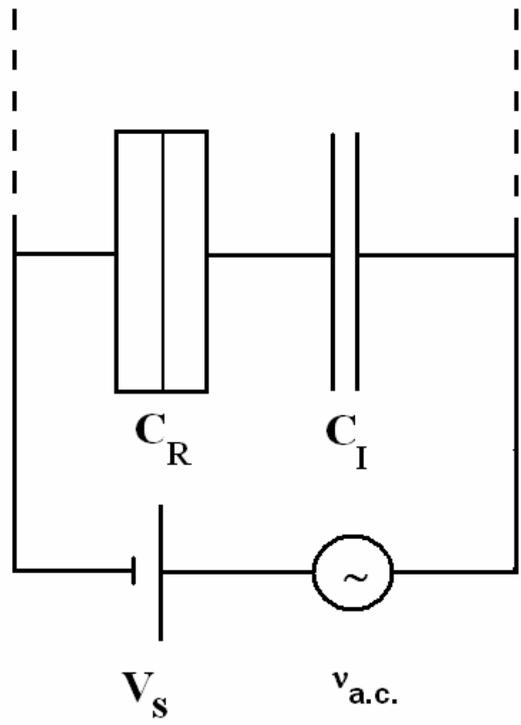

**Figure 1.A**

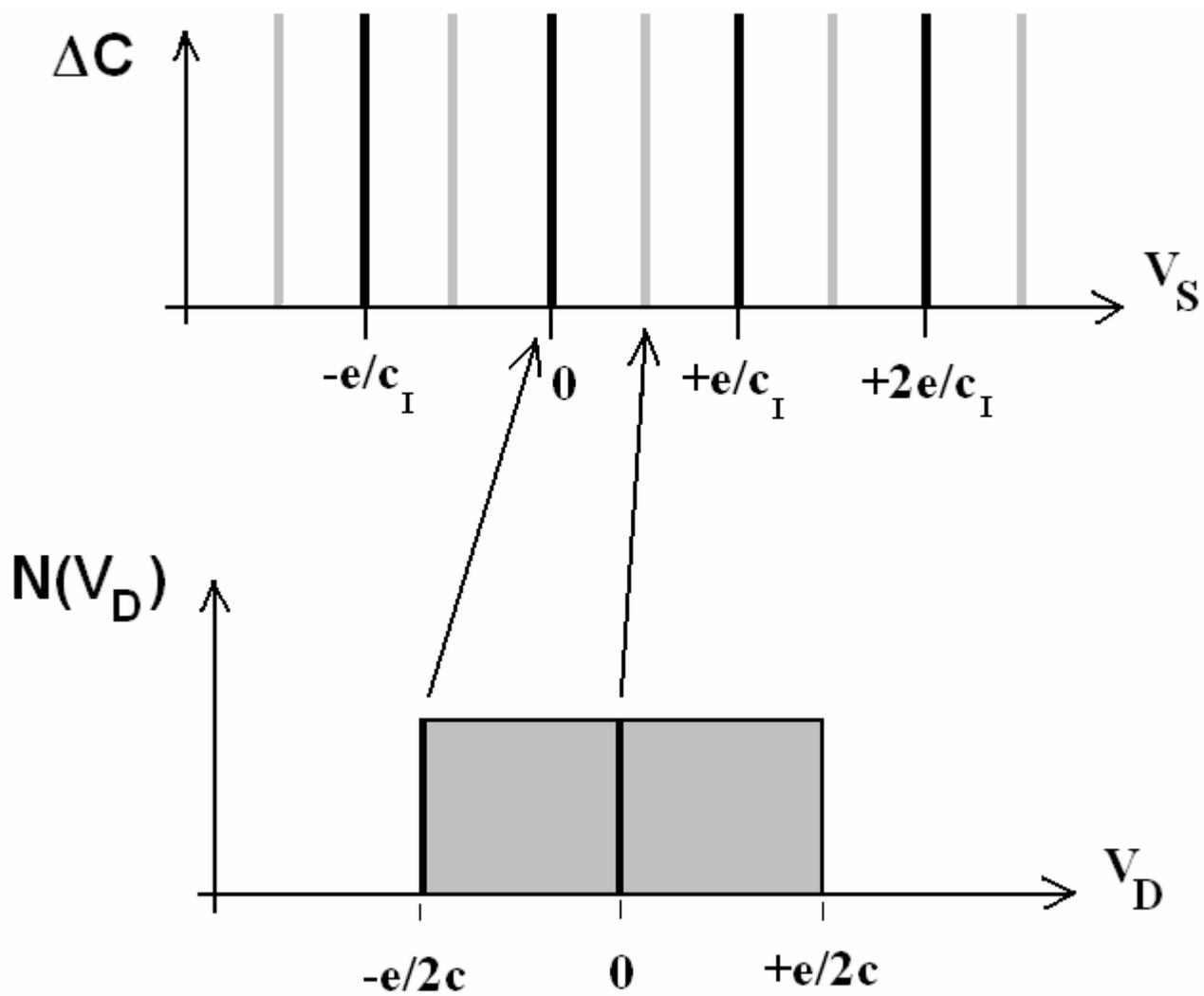

**Figure 2.A**

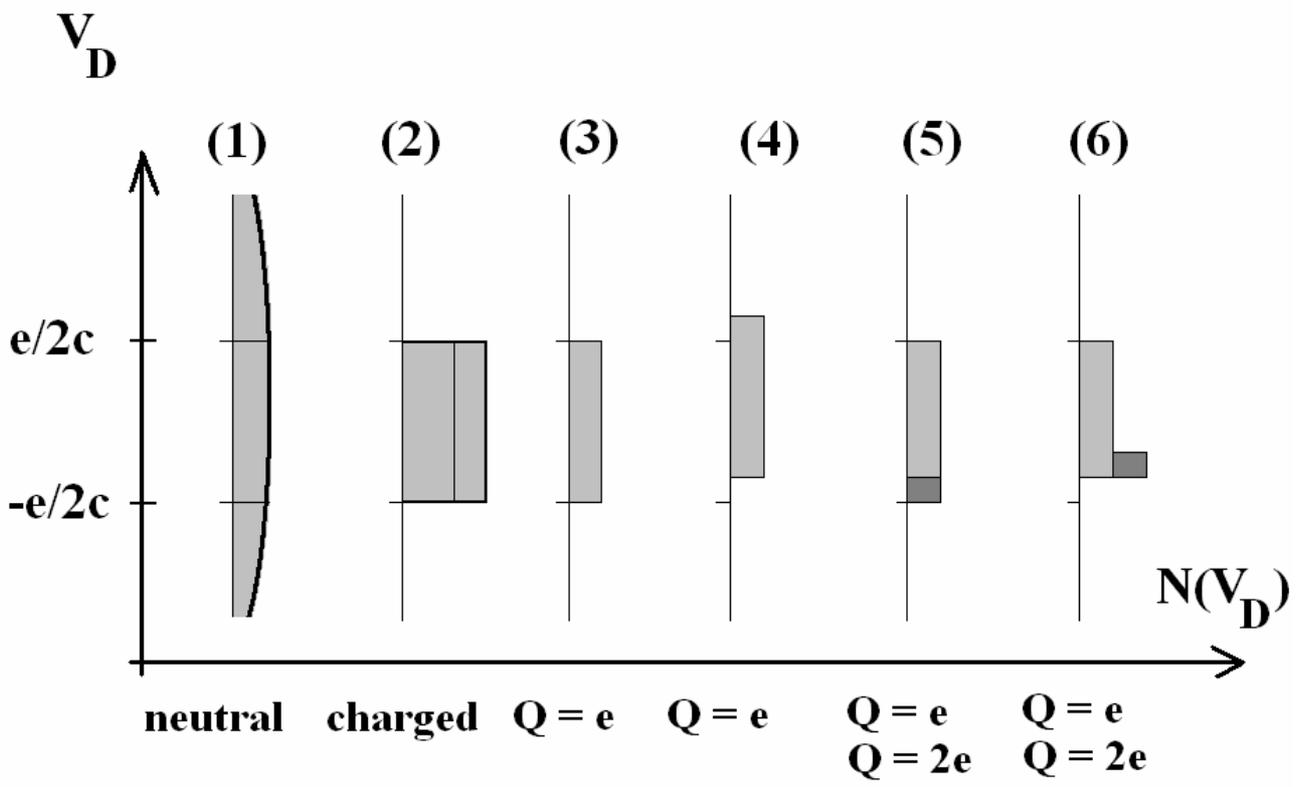

**Figure 3.A**